\documentclass[12pt]{article}
\pdfoutput=1
\usepackage{tikz}
\usetikzlibrary{matrix,arrows,positioning,shapes,snakes,fadings,decorations.pathmorphing,
decorations.pathreplacing,automata,shadows,decorations.markings,patterns}
\usepackage[thicklines]{cancel}
\usetikzlibrary{calc,shapes.callouts,shapes.arrows}

\usepackage{mathtools}
\usepackage{mathpazo}
\usepackage{jheppub}
\usepackage{amsmath,amssymb,euscript,array,mathrsfs,appendix,ctable,marvosym}
\usepackage{arydshln}
\usepackage{todonotes}
\usepackage{graphicx}
\usepackage[normalem]{ulem}

\newcommand\blank[1]{#1}
\renewcommand\blank[1]{}
\def\Buildrel#1\over#2\under#3{\mathrel{\mathop{\kern0pt
#2}\limits^{#1}_{#3}}}

\newcommand*\circled[1]{\footnotesize\tikz[baseline=(char.base)]{%
            \node[shape=circle,fill=black!20,draw,inner sep=2pt] (char) {#1};}}

\usepackage{enumitem}

\def\Lax{{\mathscr L}}
\def\CF{{\cal F}}

\def\SO{\text{SO}}
\def\JJ{\mathscr{J}}

\def\msl{\mathfrak {sl}}
\def\msu{\mathfrak {su}}

\def\msu{\mathfrak{su}}

\def\msl{\mathfrak{sl}}

\def\a{\alpha}
\def\b{\beta}

\newcommand{\csch}{\operatorname{csch}}
\newcommand{\Tr}{\operatorname{Tr}}
\newcommand{\cn}{\operatorname{cn}}
\newcommand{\sn}{\operatorname{sn}}
\newcommand{\dn}{\operatorname{dn}}

\newcommand{\IM}{\operatorname{Im}}

\def\B0{{\boldsymbol 0}}

\def\BId{{\boldsymbol{I}}}
\def\BR{{\boldsymbol{\cal R}}}
\def\BOmega{{\boldsymbol\Omega}}

\def\BTheta{{\boldsymbol\Theta}}

\def\det{{\rm det}}
\def\SU{\text{SU}}
\def\U{\text{U}}

\def\SL{\text{SL}}

\def\ee{\boldsymbol{e}}

\newcommand{\Be}{\boldsymbol{e}}

\def\Dbarslash{\,\,{\raise.15ex\hbox{/}\mkern-12mu {\bar D}}}
\def\Dslash{\,\,{\raise.15ex\hbox{/}\mkern-12mu D}}
\def\delslash{\,\,{\raise.15ex\hbox{/}\mkern-9mu \partial}}
\def\delbarslash{\,\,{\raise.15ex\hbox{/}\mkern-9mu {\bar\partial}}}

\def\ket#1{| #1\rangle}

\def\LAG{\mathscr{L}}

\newcommand{\MAT}[1]{\begin{pmatrix} #1\end{pmatrix}}

\newcommand{\EQ}[1]{\begin{equation}\begin{split} #1
\end{split}\end{equation}}

\def\be{\begin{equation}}
\def\ee{\end{equation}}

\title{Yang Baxter and Anisotropic Sigma and Lambda Models, Cyclic RG and Exact S-Matrices}
\author{Calan Appadu, Timothy J. Hollowood, Dafydd Price and Daniel C. Thompson}
\affiliation{Department of Physics, Swansea University, Swansea, SA2 8PP, U.K.}

\emailAdd{t.hollowood@swansea.ac.uk}
\emailAdd{D.C.Thompson@Swansea.ac.uk}

\abstract{Integrable deformation of $\SU(2)$ sigma and lambda models are considered at the classical and quantum levels. These are the Yang-Baxter and XXZ-type anisotropic deformations. The XXZ type deformations are UV safe in one regime, while in another regime, like the Yang-Baxter deformations, they exhibit cyclic RG behaviour. The associated affine quantum group symmetry, realized classically at the Poisson bracket level,  has $q$ a complex phase in the UV safe regime and $q$ real in the cyclic RG regime, where $q$ is an RG invariant.
Based on the symmetries and RG flow we propose exact factorizable S-matrices to describe the scattering of states in the lambda models, from which the sigma models follow by taking a limit and non-abelian T-duality. In the cyclic RG regimes, the S-matrices are periodic functions of rapidity, at large rapidity, and in the Yang-Baxter case violate parity.}

\notoc
\begin{document}

\pgfdeclarelayer{background layer} 
\pgfdeclarelayer{foreground layer} 
\pgfsetlayers{background layer,main,foreground layer}

\maketitle

\newpage

\section{Introduction}\label{s1}

Sigma models are fascinating because they are the building blocks of string worldsheet theories but also they share many of the features of QFTs in higher dimensions in a simpler context. And within the space of sigma models, the ones that are integrable have the additional lure of tractability. 

The key examples are the Principal Chiral Models (PCM), whose target spaces are group manifolds $G$. There is a $G$-valued field $f$ and the action can be written\footnote{We take $x^\pm= t\pm x$ and so for vectors $A^\pm=A^0\pm A^1$ along with $A_\pm=\frac12(A^0\pm A^1)$.}
\EQ{\label{pcm2}
S=-\frac 1{2\pi\alpha}\int d^2x\,\Tr\big[f^{-1}\partial_+f\,f^{-1}\partial_-f\big]\ .
}
The PCM  can appear as a bosonic sub-sector of a consistent string theory CFT background, e.g.~the $D1$-$D5$ near horizon geometry, providing a modern holographic motivation for studying this theory. The more prosaic view, which we adopt here, is that PCMs are an exceptionally informative $1+1$-dimensional QFTs exhibiting asymptotic freedom in the running coupling $\alpha(\mu)$ and a dynamically generated mass gap.  

The action given in Eq.~\eqref{pcm2}  manifests a $G_L\times G_R$ global symmetry, $f\to UfV$.  A feature that makes the PCM tractable is that it is classically integrable and the  $G_L\times G_R$ symmetry is part of a much larger classical Yangian  ${\mathscr Y}(\mathfrak g_L) \times {\mathscr Y}(\mathfrak g_R)$ symmetry generated by non-local charges.\footnote{A concise introduction to this symmetry can be found in \cite{MacKay:2004tc}. There are also an infinite number of local conserved charges which include and energy and momentum.}  

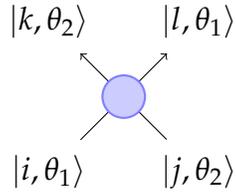
\begin{figure}
\begin{center}
\begin{tikzpicture} [line width=1.5pt,inner sep=2mm,
place/.style={circle,draw=blue!50,fill=blue!20,thick}]
\begin{pgfonlayer}{foreground layer}
\node at (1,1) [place] (sm) {}; 
\end{pgfonlayer}
\node at (0,0) (i1) {$|i,\theta_1\rangle$};
\node at (2,0) (i2) {$|j,\theta_2\rangle$};
\node at (0,2) (i3) {$|k,\theta_2\rangle$};
\node at (2,2) (i4) {$|l,\theta_1\rangle$};
\draw[->] (i1) -- (i4);
\draw[->] (i2) -- (i3);
\end{tikzpicture}
\caption{\small The basic $2\to2$ S-matrix elements depend on the rapidities of the incoming and outgoing particles as well as the internal quantum numbers $i,j,k,l$.}
\label{f1} 
\end{center}
\end{figure}

At the quantum level this integrability persists leading to the  factorization of its S-matrix  \cite{Luscher:1977uq,Goldschmidt:1980wq}. This means that it is completely determined by $2\to2$ body processes which preserve the individual momenta, as illustrated in Fig.~\ref{f1}.
The states are labelled by their rapidity $\theta$ and by internal quantum numbers $i,j,\ldots$. For example, in the $\SU(N)$ PCM, there are $N-1$ particle multiplets with mass $m_a=m\sin(\pi a/N)$, $a=,1,2,\ldots,N-1$, and each multiplet transforms in the $[\omega_a]\times[\omega_a]$ representation of the $G_L\times G_R$ symmetry, where $\omega_a$ are the highest weight vectors of the $a^\text{th}$ fundamental representation.\footnote{For the groups $\SO(N)$ the representations are actually reducible combinations.} The 2-body S-matrix has the characteristic product form \cite{Ogievetsky:1987vv}:
\EQ{
S(\theta)=S_{G_L}(\theta)\otimes S_{G_R}(\theta)\ ,
\label{glr}
}
where $\theta=\theta_1-\theta_2$. The product form
reflects the fact that the states transform in a product of representations of $G_L$ and $G_R$. The S-matrix building block $S_G(\theta)$ is $G$-invariant, in fact Yangian invariant, and is built from a rational solution of the Yang-Baxter Equation.\footnote{For the higher rank groups, the product form of the S-matrix must be multiplied by a scalar factor to provide the bound state poles.}

Since the PCM is asymptotically free and its spectrum is massive and dynamically generated, directly connecting the conjectured quantum S-matrix picture to the Lagrangian description in Eq.~\eqref{pcm2} is subtle.  Nonetheless, consistency checks can be made by studying the theory in a regime in which perturbation theory can be employed and compared against the factorized S-matrix.  The study of the the exact solution of the model was initiated in the classic works \cite{Polyakov:1983tt,Wiegmann:1984ec,Wiegmann:1984pw,Ogievetsky:1987vv}.  As a byproduct of the successful comparison of Thermodynamic Bethe Ansatz and perturbative calculations of the free energy in a background charge one obtains an exact expression for the mass gap.\footnote{The case of $G=\SU(2)$ viewed as the $O(4)$ $\sigma$-model was done in \cite{Hasenfratz:1990zz},  with the extension to $\SU(N)$ in \cite{Balog:1992cm,Fateev:1994ai} and other Lie algebras in \cite{Hollowood:1994np}.}

A natural question to ask, is whether the PCM can be deformed in a way that preserves integrability? For the case $\SU(2)$---which we we will concentrate on in this work---there are several ways to do this, while for higher rank groups the possibilities appear to be more limited.  We will concentrate on the deformations that preserve one of the chiral symmetries, $\SU(2)_L$, say. Deformation which preserve the $\SU(2)_L$ symmetry can be written
\EQ{\label{eq:defpcm}
S=-\frac1{2\pi}\int d^2x\,\Tr\big[f^{-1}\partial_+f \BTheta f^{-1}\partial_-f\big]\ ,
}
where $\BTheta$ is endomorphism of the Lie algebra, $\BTheta\cdot T^a=\BTheta_{ab}T^b$.  A fascinating problem is to determine systematically which choices of $\BTheta$ lead to integrable models both classically and quantum mechanically.   

For the particular case of $G=\SU(2)$, there are {\it anisoptropic\/} type deformations that involve in the most general case three different couplings $\BTheta\cdot T^a=\alpha_a^{-1}T^a$. Introducing the components of the $\SU(2)_L$ current $J_\mu=\sum_aJ^a_\mu T^a$,\footnote{Throughout the paper we use a basis $\{T^a\}$ that are anti-hermitian and normalized so that $\Tr(T^aT^b)=-\delta^{ab}$. So for $\SU(2)$, $T^a=i\sigma^a/\sqrt2$ where $\sigma^a$ are the Pauli matrices. In addition, we define $T^\pm=(T^1\pm iT^2)/\sqrt2$ and the alternative decomposition $J_\mu=J_\mu^3T^3+J_\mu^+T^-+J_\mu^-T^+$.} we can write the action for the most general deformation of this type as  \cite{Wiegmann:1985jt,Faddeev:1985qu}
\EQ{
S=\frac1{2\pi}\int d^2x\,\Big[\frac1{\alpha_1}J_+^1J_-^1+\frac1{\alpha_2}J_+^2J_-^2+\frac1{\alpha_3}J_+^3J_-^3\Big]\ .
}
We denote these kinds of deformations as the XYZ or XXZ type, depending upon whether the $\alpha_i$ are all different or two are equal, respectively. Surprisingly, these kinds of deformations are special for $\SU(2)$ and generalizations of this type fail to be integrable for higher rank groups.

The key to generalizing integrable deformations to arbitrary groups was uncovered some years ago by Klimcik \cite{Klimcik:2002zj,Klimcik:2008eq}. These are the  Yang-Baxter (YB) deformations of the PCM that are associated to $\BR$, an antisymmetric endomorphism of the Lie algebra that satisfies the (modified) classical YB equation
\EQ{
[\BR a,\BR b]-\BR [ a,b]_\BR =-c^2[a,b]\ , 
\label{yb1}
}
where we have defined the $\BR$-Lie bracket
\EQ{
[ a,b]_\BR =
[\BR a,b]+[a,\BR b] \ , 
\label{yb1}
}
for all $a,b$ in the Lie algebra and where $c$ is a free parameter. The action of the deformed theory is defined by taking in \eqref{eq:defpcm}
\EQ{
\BTheta=\alpha^{-1}(1-\eta\BR )^{-1}\ ,
}
where $\eta$ is the real deformation parameter.   YB deformations of this type can be defined for an arbitrary group and in general the deformed theories have a Kalb-Ramond field which correspond to the terms odd in $\BR $ when the operator $(1-\eta\BR )^{-1}$ is expanded in powers of $\BR $. 

For $\SU(2)$ there is a single class of deformations of this type which, without loss of generality, can be written as
\EQ{
\BR _{ab}=\MAT{0 & -1 & 0\\ 1 & 0 & 0\\ 0 & 0 & 0}\  .
}
This satisfies \eqref{yb1} with $c^2=-1$. In this case, one can show that the Kalb-Ramond field is a total derivative and---at least with periodic boundary conditions---the YB and XXZ sigma theories are equivalent with
\EQ{\label{eq:abtoetat}
1+\eta^2= \frac\beta\alpha\ ,
}
in the regime with $\beta>\alpha$.

On a group manifold one can go further and define an integrable two-parameter ``bi-Yang-Baxter'' deformation  \cite{Klimcik:2014bta} with 
\EQ{
\BTheta=\alpha^{-1} (1-\eta\BR - \zeta \BR_f )^{-1}\ ,
}
in which $ \BR_f = {\textrm{Ad}}_f  \cdot \BR  \cdot  {\textrm{Ad}}_{f^{-1}} $.   Notice that $\BTheta$ now depends on the group element $f$ and consequently this deformation breaks both left and right acting global symmetries.  The whole construction works for an arbitrary Lie group, but specialised to $\SU(2)$ the Kalb-Ramond two-form is pure gauge and it was shown in \cite{Hoare:2014pna} that this theory matches the full two parameter Fateev model  \cite{Fateev:1996ea}.\footnote{The matching of parameters (defined after Eq. (76) of  \cite{Fateev:1996ea}) is given by  \begin{equation*}
\eta^2=\frac{r}{u}(\ell u^{-1} +1) \ , \quad
\zeta^2=\frac{\ell}{u} (r u^{-1} +1) \ ,\quad
\alpha=u\ .
 \end{equation*} } We will not consider this more general deformation any further and focus on deformations that preserve the $\SU(2)_L$ symmetry because these cases have an associated lambda model.
 
The lambda models are a completely different class of integrable deformations of the PCM. In fact of each of the sigma models, whether PCM, XXZ, XYZ or YB, i.e.~all having an $\SU(2)_L$ symmetry, have an associated lambda model that inherits the integrability of the parent sigma model.
Motivated by the process of non-abelian T-duality in string theory, each sigma model whose target space is a $G$ group manifold with $G_L$ global symmetry has an associated lambda model.\footnote{There are also examples associated to symmetric space quotients $G/H$ that we will not consider here.}  The definition of the lambda model associated to the $\SU(2)$ PCM go back to \cite{Evans:1994hi} but in a more general context are best constructed by Sfetsos's gauging procedure \cite{Sfetsos:2013wia}: 
\begin{enumerate}
\item Write down a theory which is the sum of the actions of the sigma model Eq.~\eqref{eq:defpcm} and a WZW model for a $G$-valued field $\CF$.
\item Gauge the joint $G$ symmetry, which acts on the WZW field by vector action $\CF\to U\CF U^{-1}$ and the sigma model field by left action $f\to Uf$.
\item Gauge fix the $G$ symmetry by setting the sigma model field $f=1$. 
\end{enumerate}

Applied to  the deformed PCM defined in Eq.~\eqref{eq:defpcm}, the result of this procedure leads to a deformation of a $G$ WZW model written in the following way:
\EQ{
S=k\, S_\text{gWZW}[\CF,A_\mu]-\frac1{2\pi}\int d^2x\,\Tr\big[A_+\BTheta A_-\big]\ ,
\label{sma}
}
where $A_\mu$ is the original $G$-valued gauge field which now plays the role of a non-propagating auxiliary Gaussian field that can be integrated out. The first term is the gauged WZW model action \cite{Witten:1983ar,Bardakci:1987ee,Gawedzki:1988nj,Karabali:1988au,Karabali:1989dk}  for a $G$-valued field $\CF$, where the whole vector $G$ symmetry is gauged, and $k\in\mathbb Z$ is the level. What is crucial for us is that if the original sigma model is integrable then so is the associated lambda model. There is also a sense that the original sigma model is recovered in the limit $k\to\infty$ along with a non-abelian T-duality \cite{Sfetsos:2013wia}. It is noteworthy that this relation is also seen  quantum mechanically at the level of the S-matrix where non-abelian T-duality manifests as an IRF-to-vertex transformation on the space of asymptotic states \cite{Appadu:2017fff}.\footnote{It is worth remarking that at the classical level non-abelian T duality can be thought of as a canonical transformation \cite{Lozano:1995jx} while at the quantum level the the IRF-to-vertex transformation can be thought of as a change of basis in the Hilbert space \cite{LeClair:1989wy,Bernard:1990ys,Bernard:1990cw}. It would be interesting to make the connection between the two phenomena more explicit.}

A fascinating question is to understand whether these integrable deformations persist in the quantum theory and if so, what are their factorizable S-matrices. We have already remarked that the PCM S-matrix takes the product form of two rational factors \eqref{glr} that manifest the Yangian $\mathscr Y(\msu(2)_L)\times\mathscr Y(\msu(2)_R)$ symmetry. This form seems to generalize: the XXZ models in the regime $\beta<\alpha$ lie in the class of ``SS models" considered by Fateev \cite{Fateev:1996ea},\footnote{In terms of Fateev's more general model with $\U(1)\times\U(1)$ symmetry and parameters $(a,b,c,d)$, we have $a^2=u(u+\ell)$, $b=0$ and $c=d=\ell/2$ and $\gamma_\perp=(u+\ell)^{-1}$ and $\gamma_3=u^{-1}$. Then
\EQ{
\gamma_\perp=(\pi\beta)^{-1}\ ,\qquad \gamma_3=(\pi\alpha)^{-1}\  , \qquad \gamma_3<\gamma_\perp \ .
}}
which have an S-matrix of the form \cite{Wiegmann:1985jt}
\EQ{
S_{\sigma\text{-XXZ}}(\theta)=S_{\SU(2)_L}(\theta)\otimes S(\theta;\gamma')\ .
\label{w8e}
}
In this expression, $S(\theta;\gamma')$ is the S-matrix of the sine-Gordon theory with coupling\footnote{Our $\gamma'$ is $\gamma'/8\pi$ of Zamolodchikov and Zamolodchikov \cite{Zamolodchikov:1978xm}. For us the breather spectrum is $m_n=2M\sin(\pi n\gamma'/2)$, $n=1,2,\ldots<\gamma^{\prime-1}$.}
\EQ{
\frac{\beta^2}{8\pi}=\frac{\gamma'}{1+\gamma'}\ .
}

The tensor product form of the S-matrix in \eqref{w8e} will prove ubiquitous and deserves some comment. Like the PCM S-matrix \eqref{glr} it reflects the factor that the particle states carry two sets of quantum numbers which under scattering are completely independent.

The XXZ deformation has broken the $\SU(2)_R$ Yangian symmetry but rather than disappearing it is deformed to an affine quantum group $\mathscr U_q(\widehat{\msu(2)})$ symmetry, where the deformation parameter
\EQ{
q=\exp\big[-i\pi/\gamma'\big]\ .
}
The parameter $\gamma'$ is an RG invariant combination of the couplings $\alpha$ and $\beta$ to be described in section \ref{s3}. Note that for $\gamma'<1$, the model has bound states that correspond to the breathers of the sine-Gordon theory. In the present context, the $n^\text{th}$ breather transforms as a singlet under $\mathscr U_q(\msu(2))$ but as a reducible representation of $\SU(2)_L$ corresponding to the tensor product of $n$ spin $\frac12$ representations.

The XXZ model of Fateev displays an important general feature of the integrable deformations: Yangian symmetries generally get  deformed into affine quantum group symmetries. The label ``quantum" here might be thought a misnomer because the quantum group symmetries are manifest in the classical theory at the Poisson bracket level \cite{Kawaguchi:2011pf,Kawaguchi:2012ve,Kawaguchi:2012gp}. This point deserves some comment. We shall show that the deformation parameter $q$ does indeed depend on $\hbar$ (or more precisely the coupling that plays the role of $\hbar$) as $q=\exp[\zeta\hbar]$. However, there is a consistent classical limit, where $\hbar\to0$ but the coupling constant dependent quantity $\zeta\to\infty$ such that $q$ is fixed. In addition, as part of the overall consistency we will show that the $q$ is an Renormalization Group (RG) invariant and so the quantum group symmetries are well defined in the quantum theory and the classical limit where it becomes realized at the Poisson bracket level.

The lambda deformations also have a characteristic effect on the S-matrix \cite{Appadu:2017fff}. For the PCM itself, the deformation changes the S-matrix block for the $\SU(2)_L$ symmetry into an affine quantum group invariant block, but realized in the Interaction-Round-a-Face (IRF), or Restricted-Solid-On-Solid (RSOS), form:\footnote{This type of S-matrix block appears in the context of the ``restricted sine-Gordon theory'" \cite{LeClair:1989wy,Bernard:1990ys,Bernard:1990cw} and also perturbed WZW models \cite{Ahn:1990gn}.} 
\EQ{
S_{\lambda\text{-PCM}}(\theta)=S_\text{RSOS}(\theta;k)\otimes S_{\SU(2)_R}(\theta)\ .
}
The new RSOS S-matrix piece implies that the states carry kink quantum numbers and the quantum group deformation parameter is a root of unity $q=\exp[-i\pi/(k+2)]$.   In contrast to the quantum group symmetry exhibited by the XXZ model, here this really is a quantum feature; in the classical regime the deformation parameter tends to unity. 

The original PCM S-matrix is recovered in the limit $k\to\infty$, where the kink factor becomes unrestricted, and then an IRF-to-vertex transformation which is the S-matrix manifestation of non-abelian T-duality:
\EQ{
S_\text{RSOS}(\theta;k)\overset{k\to\infty}{\xrightarrow{\hspace*{1.3cm}}} S_\text{SOS}(\theta) \overset{\text{IFR-to-vertex}}{\xrightarrow{\hspace*{1.3cm}}}\ S_\text{\SU(2)}(\theta)\ .
}

It is tempting to think that the S-matrix product form---the SS form of Fateev---will describe all the integrable deformations of the PCM and this intuition will turn out to be true.
In this paper, we will concentrate on the XXZ and YB deformations of the $\SU(2)$ PCM and their associated lambda models at the quantum level and map out their renormalization structure and their S-matrices and confirm the ubiquity of the product form. 
Specifically in this paper we:
\begin{enumerate}[label=\protect\circled{\arabic*}]
\item Review the classical integrability of the deformed sigma models and establish some new results for the Poisson brackets of the associated lambda models.
\item We show that the lambda models have quantum group symmetries in the classical theory realized at the level of the Poisson brackets. 
\item We then consider the RG flow of the sigma and lambda models at one loop order (so in the lambda models to leading order in $1/k$). We show that the XXZ models, both sigma and lambda, have one regime which has UV safe flows, whereas in the other regime there are cyclic RG type flows. The YB lambda model also has cyclic RG flows. 
\item We show that the quantum deformation parameters $q$ of the classically-realized quantum groups are RG invariants.
\item Using the RG flow and the structure of the classical symmetries, we propose S-matrices to describe all the lambda models. For the examples with cyclic RG flow, the S-matrix has periodicity in the rapidity when the rapidity is large.
\item We then argue that S-matrices of the sigma models are obtained in the large $k$ limit after an IRF-to-vertex transformation. 
\end{enumerate}

In a follow up paper, we will address the question of whether the theories that we find with cyclic RG behaviour actually exists as QFTs in the the continuum limit \cite{forthcoming}. We will find that the continuum theories can be formulated as a Heisenberg XXZ spin chain. When the RG flow of the theory has a UV safe limit, the spin chain is critical and a continuum limit can be defined. On the contrary in the regime with cyclic RG flows, the spin chain has a gap and a continuum limit does not exist. Th conclusion would be that the theories with cyclic RG behaviour only exist as effective theories with an explicit cut off.
 
\section{Classical $\SU\boldsymbol{(2)}$ Sigma Models}\label{s2}

In this section, we consider some of the aspects of the sigma models and in particular the symmetries, that will inform our S-matrix hypotheses.

\subsection{Lax connection and Poisson brackets}\label{s2.1}

The most direct way to prove classical integrability is to write down the equations of motion in Lax form, that is as the flatness condition on an auxiliary connection that depends on an additional free parameter, the {\it spectral parameter\/},
\EQ{
[\partial_++\LAG_+(z),\partial_-+\LAG_-(z)]=0\ ,
}
for arbitrary $z$.  

If we define the $\SU(2)_L$ invariant current $J_\mu=f^{-1}\partial_\mu f=J^a_\mu T^a$, the equations of motion along with the Cartan-Maurer identity of the YB deformed sigma models can be written in Lax form with
\be\label{eq:laxYB}
\LAG_\pm(z) =  \left( \frac{z \pm \eta^2}{z \mp 1} \pm  \eta \BR    \right)({\bf 1} \pm \eta \BR  )^{-1} J_\pm \ .
\ee
There are alternative ways of writing the Lax connection which differ from the above by a gauge transformation \cite{Kawaguchi:2012gp,Kawaguchi:2012ve,Delduc:2017brb}. We note in passing that the Lax connection is valued in the loop algebra $\widehat{\msu(2)}=\msu(2)\otimes\mathbb C[z,z^{-1}]$, the untwisted affinization of $\msu(2)$ (with vanishing centre). This can also be described as the affine algebra with the {\it homogeneous\/} gradation and we will denote it as $\widehat{\msu(2)}_h$.

For the anisotropic models, the Lax connection take a characteristic form that generalizes nicely as one goes through the hierarchy from PCM to XXZ to XYZ:
\EQ{
\Lax_\pm(z)=\sum_{a=1}^3w_a(\nu\mp z)J^a_\pm T^a\ .
\label{h5e}
}

For the PCM, the functions $w_a(z)$ are rational
\EQ{
w_a(z)=\frac\nu z\ ,
}
while for the XXZ case, with $\alpha_a=(\beta,\beta,\alpha)$, the functions $w_a(z)$ are trigonometric (or hyperbolic)  \cite{Faddeev:1985qu},
\EQ{
w_1(z)=w_2(z)=\frac{\sinh\nu}{\sinh z}\ ,\qquad w_3(z)=\frac{\tanh\nu}{\tanh z} \ ,
}
where 
\EQ{
\cosh^2\nu=\frac\alpha\beta\ .
}
For these theories, if we transform to a multiplicative spectral parameter $z\to\log z$ and then expand in powers of $z$, it is noteworthy that the Lax connection takes values in the {\it twisted\/} loop algebra, where the twist is an automorphism $\tau$:
\EQ{
\tau(T^{1,2})\to-T^{1,2}\ ,\qquad \tau(T^3)\to T^3\ .
}
 The Lie algebra splits into its eigen-spaces under $\tau$ and in the  {\it twisted\/} loop algebra each eigen-space receives a different scaling of the spectral parameter.   The twisted loop algebra thus has elements $T^3z^{2n}$, $T^1z^{2n+1}$ and $T^2z^{2n+1}$, with $n\in\mathbb Z$. Since the automorphism $\tau$ is {\it inner\/} the twisted loop algebra is simply equal to original in another gradation, in this case it called the {\it principal\/} gradation and we denote it $\widehat{\msu(2)}_p$.

Finally, for the XYZ case, the functions $w_a(z)$ are elliptic functions
\EQ{
w_1(z)&=i\sqrt{\frac{\alpha_2}{\alpha_1-\alpha_2}}\cdot\frac1{\sn z}\ ,\\
w_2(z)&=i\sqrt{\frac{\alpha_1}{\alpha_1-\alpha_2}}\cdot\frac{\cn z}{\sn z}\ ,\\
w_3(z)&=i\sqrt{\frac{\alpha_1\alpha_2}{\alpha_3(\alpha_1-\alpha_2)}}\cdot\frac{\dn z}{\sn z}\ ,
}
where the Jacobi elliptic functions have an elliptic modulus
\EQ{
k^2=\frac{\alpha_1-\alpha_3}{\alpha_1-\alpha_2}\ .
}
In addition,
\EQ{
\cn^2\nu=\frac{\alpha_2}{\alpha_1}\ .
}
Whilst these theories still have the $\SU(2)_L$ symmetry, the $\SU(2)_R$ symmetry is broken to a finite $\mathbb Z_4$ subgroup. The question as to whether the associated Yangian symmetry becomes deformed is an interesting one that we do not tackle here. Note that since the $\SU(2)_L$ symmetry is preserved these theories are distinct from the general two-parameter deformations considered in \cite{Fateev:1996ea}. 

Note that the XXZ model in the regime $\alpha>\beta$ has the same equation of motion as the YB model but the Lax connections are completely different. The relation between the two formulations was considered in detail in \cite{Kawaguchi:2012gp}.

As part of the standard formalism of integrability (e.g.~see the book \cite{BBT}), a key structure is the Poisson bracket of the spatial component of the Lax connection $\Lax\equiv\Lax_+-\Lax_-$. This is sometimes called the Maillet algebra  \cite{Maillet:1985ek} and in general  takes the form
\EQ{
&\{\Lax_1(x;z),\Lax_2(y;w)\}=[r(z,w),\Lax_1(x;z)+\Lax_2(x;w)]\delta(x-y)\\ &\qquad\qquad-[s(z,w),\Lax_1(x;z)-\Lax_2(y;w)]\delta(x-y)-2s(z,w)\delta'(x-y)\ .
\label{ikk}
}
The notation means that the bracket acts on a product of $\msu(2)$ modules $V\otimes V$ and the subscripts indicate which of the copies a quantity acts on: $\Lax_1(z)=\Lax(z)\otimes 1$ and $\Lax_2(z)=1\otimes\Lax(z)$. The tensor kernels $r(z,w)$ and $s(z,w)$ act on $V\otimes V$. 

In many cases, the kernels $r$ and $s$ can be written in the form
\EQ{\label{eq:rs}
r(z,w)&=\frac{\phi(w)^{-1}+\phi(z)^{-1}}{z-w}\Pi\ ,\\ 
s(z,w)&=\frac{\phi(w)^{-1}-\phi(z)^{-1}}{z-w}\Pi\ ,
}
where $\phi(z)$ is known as the {\it twist function\/} and in many cases $\Pi=-\sum_aT^a\otimes T^a$ is the Casimir tensor. For example, for the YB deformation with the definition of the Lax connection in  \cite{Kawaguchi:2012gp}, the kernels $r$ and $s$ take precisely this form with a twist function
\EQ{
\phi(z)=\frac1{2\pi\beta}\cdot\frac{1-z^2}{\eta^2+z^2} \ .
\label{t4r}
}
Note that here we include the factor $2\pi\beta$ which plays the role of $\hbar$ in the quantum theory. 

For the XXZ model in the trigonometric formulation, the $r/s$ kernels take a similar form, except that 
\EQ{
\Pi\equiv\Pi(z,w)=-T^1\otimes T^1-T^2\otimes T^2-\cosh(z-w)T^3\otimes T^3\ ,
\label{dpi}
}
depends on $z$ and $w$, and the twist function
\EQ{
\phi(z)=\frac1{2\pi\beta\sqrt{\alpha^2-\alpha\beta}}\cdot\frac{\alpha-\beta\cosh^2z}{\sinh^2z}\ .
\label{s8i}
}

\subsection{Non-local charges and infinite symmetries}\label{s2.2}

Integrable field theories have an infinite sets of both local (integrals of expressions local in the fields and their derivatives) and non-local conserved charges. All these charges can be extracted from the Lax connection. The local conserved charges include the energy and momentum but the non-local ones are our central focus here because they generate some remarkable infinite symmetries in the form of Yangians and quantum groups. 

The non-local charges are encoded in the monodromy matrix, the parallel transport of the Lax connection, along the spatial direction
\EQ{
T(z)=\text{P}\overset{\longleftarrow}{\text{exp}}\Big[-\int_{-\infty}^\infty dx\,\Lax(x;z)\Big]
}
which is conserved in time (in the infinite volume limit with appropriate fall off assumed). We can think of $T(z)$ as a generating function for the charges. It is natural to lift the Poisson bracket on $\Lax(x;z)$ to the monodromy matrix. However, this is where a problem arises as a result of the {\it non ultra-locality\/} of the Poisson bracket: when the kernel $s$ is non-trivial the Poisson bracket of the monodromy matrix is ill-defined due to the $\delta'(x-y)$ term in \eqref{ikk}. This non ultra-locality can lead to ordering ambiguities when considering nested integrals in the expansion of the monodromy matrix and a violation of the Jacobi identity for the monodromy matrix.  One way to deal with the ambiguities is to use Maillet's prescription \cite{Maillet:1985ek}. This corresponds to lifting the Poisson bracket to the monodromy matrix in the form
\EQ{
\{T_1(z),T_2(w)\}&=[r(z,w),T_1(z)T_2(w)]\\ &+T_1(z)s(z,w)T_2(w)-T_2(w)s(z,w)T_1(w)\ .
}
It is remarkable that the non ultra-locality and its associated ambiguities generally turn out not to affect the discussion of the Yangian and quantum group symmetries when they are manifested at the classical level  \cite{Delduc:2017brb}. As we will see, there can also be quantum group symmetries that can only be seen consistently at the quantum level.

The infinite symmetries are associated to the expansion of the monodromy matrix $T(z)$ around special points $z_*$ which define non-local charges that generate Yangian or quantum group symmetries. The general idea is as follows: generically the kernel $r$ has a pole as $z\to w$; however, there are special points $z_*$ in the neighbourhood of which,
\EQ{
z=z_*+\epsilon\ ,\qquad w=z_*+\tilde\epsilon\ ,
}
the Poisson bracket algebra has a finite limit as $\epsilon$ and $\tilde\epsilon$ are scaled to $0$. The special points can also be at infinity in which case one takes $z=\epsilon^{-1}$ and $w=\tilde\epsilon^{-1}$. 

If the $r/s$ kernels take the form \eqref{eq:rs}, then poles of the twist function are special points (see \cite{Vicedo:2015pna} for a general analysis for these cases). For example, for the YB deformation with twist function \eqref{t4r}, there are poles at $z=\pm i\eta$, around which
\EQ{
r(z,w)&=\pm\frac{i\pi\beta\eta}{1+\eta^2}\cdot\frac{\epsilon+\tilde\epsilon}{\epsilon-\tilde\epsilon}\cdot\sum_{a=1}^3T^a\otimes T^a+{\mathscr O}(\epsilon)\ ,\\ s(z,w)&=-\frac{i\pi\beta\eta}{1+\eta^2}\cdot\sum_{a=1}^3T^a\otimes T^a+{\mathscr O}(\epsilon)\ .
\label{pe6}
}
It has been shown that the charges defined by expansing the monodromy matrix around these special points generate a classical version of an affine quantum group symmetry $\mathscr U_q(\widehat{\msu(2)})$ with a deformation parameter \cite{Kawaguchi:2012ve,Kawaguchi:2012gp,Delduc:2017brb}
\EQ{
q=\exp\Big[-2\pi\beta\cdot\frac{\eta}{1+\eta^2}\Big]=\exp\Big[-4\pi\sqrt{\alpha\beta-\alpha^2}\Big]\ .
\label{qcl}
}
Note the factor $2\pi\beta$ comes from the overall normalization of the action and plays the role of $\hbar$, and so is usually set to 1 in a classical analysis \cite{Kawaguchi:2012ve,Kawaguchi:2012gp,Delduc:2017brb}. For us, pursuing a quantum analysis, having the correct overall normalization is crucial because the correctly defined $q$ is then an RG invariant. For Yang-Baxter deformations a similar result was obtained for arbitrary groups and also symmetric space coset $\sigma$-models in a now seminal paper \cite{Delduc:2013fga}.\footnote{For group case the result of \cite{Delduc:2013fga} is that $q=\exp[-\epsilon(1-\epsilon^2)^{3/2}]$ with $\eta=\epsilon/\sqrt{1-\epsilon^2}$ which matches the above after taking into account that the overall tension has been set as $\alpha^{-1}= (1+\eta^2)^2 $. Although not present focus it would be remiss not to mention that that a Yang-Baxter deformation of the Metsaev-Tseytlin action for strings in $AdS_5 \times S^5$ was constructed in \cite{Delduc:2013qra,Delduc:2014kha}. } 

In the expansion of $T(z)$ around $z_*=\pm i\eta$, the charges are naturally are classified by the order in which they appear \cite{Kawaguchi:2012gp} (positive/negative grade for $z_*=\pm i\eta$): see Fig.~\ref{f6}.
\begin{figure}
\begin{center}
\begin{tikzpicture}[scale=0.8]
\draw[thick] (-1,-3.4) -- (-1,3.2);
\draw[thick] (1,-3.4) -- (1,3.2);
\draw[thick] (7,-3.4) -- (7,3.2);
\node at (-3,1) {\small $z_*=+i\eta$};
\node at (-3,-1) {\small $z_*=-i\eta$};
\draw[very thick,->] (-3,1.3) -- (-3,2.3);
\draw[very thick,->] (-3,-1.3) -- (-3,-2.3);
\node at (0,-3) {$\vdots$};
\node at (0,-2)  {$-2$};
\node at (0,-1) {$-1$};
\node at (0,0)  {$0$};
\node at (0,1) {$+1$};
\node at (0,2) {$+2$};
\node at (0,3) {$\vdots$};
\node at (2,3) {$\vdots$};
\node at (4,3) {$\vdots$};
\node at (6,3) {$\vdots$};
\node at (2,-3) {$\vdots$};
\node at (4,-3) {$\vdots$};
\node at (6,-3) {$\vdots$};
\node at  (2,-2) {$\frak Q_{-2}^+$};
\node at (4,-2) {$\frak Q_{-2}^3$};
\node at (6,-2) {$\frak Q_{-2}^-$};
\node at  (2,-1) {$\frak Q_{-1}^+$};
\node at (4,-1) {$\frak Q_{-1}^3$};
\node[color=red] at (6,-1) {$\tilde{\frak Q}^-$};
\node[color=red] at  (2,0) {$\frak Q^+$};
\node[color=red] at (3.3,0) {$\frak Q^3$};
\node at (3.9,-0.07) {$=$};
\node[color=blue] at (4.7,0) {$-\tilde{\frak Q}^3$};
\node[color=blue] at (6,0) {${\frak Q}^-$};
\node[color=blue] at  (2,1) {$\tilde{\frak Q}^+$};
\node at (4,1) {$\frak Q_{1}^3$};
\node at (6,1) {$\frak Q_1^-$};
\node at  (2,2) {$\frak Q_2^+$};
\node at (4,2) {$\frak Q_2^3$};
\node at (6,2) {$\frak Q_2^-$};
\end{tikzpicture}
\caption{\footnotesize The charges and their grades for the expansion of the monodromy matrix around the pair of special points $z=\pm i\eta$. The blue/red and positive/negative graded charges are associated to $\pm i\eta$, respectively. The red and blue charges generate the affine quantum group in homogenous gradation and all the other charges are obtained by repeated Poisson brackets of these charges.}
\label{f6}
\end{center}
\end{figure}

The $\U(1)_R$ charge $\frak Q^3$, local in fields, is supplemented with non-local conserved charges $\frak Q^\pm$  that obey a (classical) quantum group $\mathscr U_q(\msu(2))$ symmetry under the Poisson bracket
\be
\{\frak Q^+ ,\frak Q^- \}  = - i \frac{q^{\frak Q^3} -q^{-\frak Q^3 } }{q - q^{-1}} \ , \quad \{\frak Q^\pm ,\frak Q^3 \} = \pm i \frak Q^{\pm} \ . 
\ee
In addition to these, one obtains generators $\tilde{\frak Q}^\pm$ associated to the affine extension\footnote{\label{footnote:gradation}Recall that the affine extension $\widehat{\msu(2)}$ supplements  the Chevalley generators $\{E_1, F_1, H_0\}$ of $\msu(2)$ with an additional root and corresponding generators $\{E_0, F_0, H_0\}$ obeying the standard relations $[H_i, E_j] = a_{ij} E_j$, $[H_i ,F_j] = -a_{ij} F_j$ and $[E_i, F_j]= \delta_{ij} H_j$ together with the Serre relations.  Here the generalised Cartan matrix $a_{ij}$ has off diagonal elements equal $-2$.   $K= H_0 + H_1$ is central and in what follows we will consider modules where $K=0$, i.e.~centreless representations  for which $\widehat{\msu(2)}$ becomes the loop algebra.  Note that we will not distinguish the real form $\msl(2)$ from $\msu(2)$ where appropriate. This being the case, representations are the tensor of an $\msu(2)$ representation and functions of a variable $z$.  There is a choice, known as gradation, to be made as to the relative action in  $\msu(2)$  space and $z$-space.  In the  {\em{homogenous}} gradation is  $E_1 = T^+ $, $F_1 = T^-$, $E_0 = z^{2} T^-$, $F_0 = z^{-2} T^+$, $H_1= - H_0 = T^3$.  In the  {\em{principal}} gradation $E_1 = z T^+ $, $F_1 = z^{-1}T^-$, $E_0 = z T^-$, $F_0 = z^{-1} T^+$, $H_1= - H_0 = T^3$.  These gradations lift to the quantum group deformation $\mathscr U_q(\widehat{\msu(2)})$.
 }
 of this symmetry (the extension is centreless since $\tilde{\frak Q}^3 = -\frak Q^3$ and so the affine algebra is actually a loop algebra). There are an infinite series of higher charges, but these can be recovered by taking repeated Poisson brackets of the charges shown.
 The grading that is imposed on the algebra by the order of the expansion that the charges appear around the special points is precisely the homogeneous gradation $\widehat{\msu(2)}_h$. The other important point is that the full set of charges that generate the affine quantum group are associated to a pair of special points.
  
In the YB sigma model there is also a special point at infinity. Setting $z=\epsilon^{-1}$ and $w=\tilde\epsilon^{-1}$, the kernels have the expansion
\EQ{
r(z,w)&=-2\pi\beta\cdot\frac{\epsilon\tilde\epsilon}{\epsilon-\tilde\epsilon}\sum_{a=1}^3T^a\otimes T^a+{\mathscr O}(\epsilon^2)\ ,\\
s(z,w)&={\mathscr O}(\epsilon^3)\ .
}
In this case, the non-vanishing contribution is at ${\mathscr O}(\epsilon)$. The charges that are defined by the expansion of monodromy matrix around infinity generate an infinite Yangian symmetry $\mathscr Y(\msu(2)_L)$ that includes the global $\SU(2)_L$ symmetry.

Now we turn to the anisotropic XXZ deformed sigma model with twist function \eqref{s8i}. In this case, the infinite symmetries are associated to the pole of the twist function at $z=0$ and to the behaviour at $\pm\infty$. Before proceeding it is more convenient to transform to multiplicative spectral parameter $z\to\log z$ in which case the twist function takes the form
\EQ{
\phi(z)=\frac1{2\pi\beta\sqrt{\alpha^2-\alpha\beta}}\cdot\frac{4\alpha z^2-\beta(z^2+1)^2}{(z^2-1)^2}\ .
}
The pole is now at $z=1$, and expanding around it, we have
\EQ{
r(z,w)&=\frac{\pi\beta\alpha}{\sqrt{\alpha^2-\alpha\beta}}\cdot\frac{\epsilon^2+\tilde\epsilon^2}{\epsilon-\tilde\epsilon}\cdot\Pi+{\mathscr O}(\epsilon^2)\ ,\\
s(z,w)&=\frac{\pi\beta\alpha}{\sqrt{\alpha^2-\alpha\beta}}\cdot(\epsilon+\tilde\epsilon)\cdot\Pi+{\mathscr O}(\epsilon^2)\ ,
}
The leading behaviour here is $\mathscr O(\epsilon)$ and so it indicative of a Yangian symmetry. In fact, expanding around this pole gives the Yangian symmetry $\mathscr Y(\msu(2)_L)$ in the trigonometric formulation.

\begin{figure}
\begin{center}
\begin{tikzpicture}[scale=0.8]
\draw[thick] (-1,-3.4) -- (-1,3.2);
\draw[thick] (1,-3.4) -- (1,3.2);
\draw[thick] (7,-3.4) -- (7,3.2);
\node at (-3,1) {\small $z_*=\infty$};
\node at (-3,-1) {\small $z_*=-\infty$};
\draw[very thick,->] (-3,1.3) -- (-3,2.3);
\draw[very thick,->] (-3,-1.3) -- (-3,-2.3);
\node at (0,-3) {$\vdots$};
\node at (0,-2)  {$-2$};
\node at (0,-1) {$-1$};
\node at (0,0)  {$0$};
\node at (0,1) {$+1$};
\node at (0,2) {$+2$};
\node at (0,3) {$\vdots$};
\node at (2,3) {$\vdots$};
\node at (6,3) {$\vdots$};
\node at (2,-3) {$\vdots$};
\node at (6,-3) {$\vdots$};
%
\node at (4,-2) {$\frak Q_{-2}^3$};
%
\node[color=red] at  (2,-1) {$\tilde{\frak Q}^+$};
\node[color=red] at (6,-1) {$\frak Q^-$};
%
\node[color=red] at (3.3,0) {$\frak Q^3$};
\node at (3.9,-0.07) {$=$};
\node[color=blue] at (4.7,0) {$-\tilde{\frak Q}^3$};
%
\node[color=blue] at  (2,1) {$\frak Q^+$};
\node[color=blue] at (6,1) {$\tilde{\frak Q}^-$};
%
\node at (4,2) {$\frak Q_2^3$};
%
\end{tikzpicture}
\caption{\footnotesize The charges and their grades for the expansion of the monodromy matrix around the pair of special points $z=\pm\infty$ (or $0,\infty$ with a multiplicative spectral parameter). The blue/red and positive/negative graded charges are associated to $\pm\infty$, respectively. The red and blue charges generate the affine quantum group in principal gradation and all the other charges are obtained by repeated Poisson brackets of these charges.}
\label{f7}
\end{center}
\end{figure}
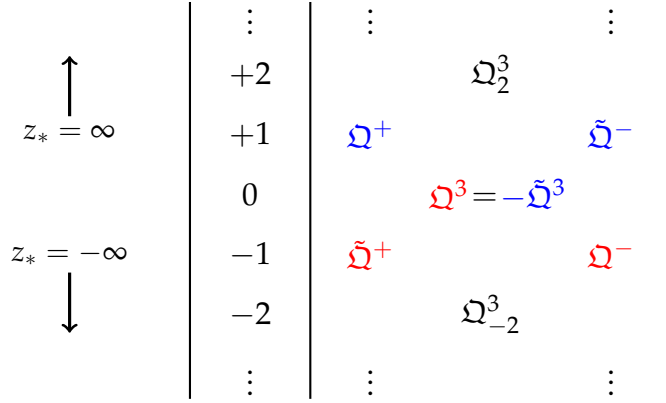

The special points at $\pm\infty$ map to $z_*=0,\infty$, around which
\EQ{
r(z,w)&=\mp2\sqrt{\alpha^2-\alpha\beta}\cdot\frac{\epsilon\tilde\epsilon}{\epsilon^2-\tilde\epsilon^2}\cdot\Pi+\mathscr O(\epsilon)\ ,\\
s(z,w)&=\mathscr O(\epsilon^2)\ .
\label{p9p}
}
The expansions in this case are associated to a quantum group symmetry with the same deformation parameter \eqref{qcl} as in the YB case, once we identify parameters as in \eqref{eq:abtoetat}.  The charges emerge as illustrated in Fig.~\ref{f7} \cite{Kawaguchi:2012gp}. Once again there are an infinite set of charges but the ones shown generate the affine algebra and the higher charges are then obtained by repeated Poisson brackets of the lower charges. The affine algebra is now revealed to be associated to the principal gradation  $\widehat{\msu(2)}_p$.

So although the YB and XXZ sigma models have the same equations of motion and what seems like identical symmetries, a Yangian and an affine  quantum group, there is a subtle difference. The affine quantum group for the YB is in the homogeneous gradation while in the XXZ case it is in the principle gradation. This interpretation is consistent with the spectral parameter rescaling of $\SU(2)$ generators found in  \cite{Kawaguchi:2012gp} required to go between the two expansion. For the YB deformation of arbitrary rank groups, for which only a rational Lax description exists, the same homogenous gradation shows itself \cite{Delduc:2017brb}.

The existence of these symmetries at the classical level is important because they will inform our search for the quantum S-matrices that describe the quantum versions of these theories. The symmetries are summarized in Table \ref{t1}.
\begin{table}
\begin{center}
\begin{tikzpicture}[scale=1]
\node at (0,0) {PCM};
\node at (0,1) {$\sigma$-YB};
\node at (0,-1) {$\sigma$-XXZ};
\node at (3,0) {$\mathscr Y(\msu(2))$};
\node at (3,1) {$\mathscr Y(\msu(2))$};
\node at (3,-1) {$\mathscr Y(\msu(2))$};
\node at (6,0) {$\mathscr Y(\msu(2))$};
\node at (6,1) {$\mathscr U_q(\widehat{\msu(2)}_h)$};
\node at (6,-1) {$\mathscr U_q(\widehat{\msu(2)}_p)$};
\draw[very thick] (-1,-1.5) -- (7.5,-1.5);
\draw[very thick] (-1,2.5) -- (7.5,2.5);
\draw[thick] (-1,1.5) -- (7.5,1.5);
\node at (0,2) {Model};
\node at (3,2) {Left symm.};
\node at (6,2) {Right symm.};
\end{tikzpicture}
\caption{\footnotesize The symmetries of the sigma models. The deformation parameter of the quantum group is given in \eqref{qcl} in terms of the underlying coupling constants. The only (subtle) difference between the symmetries is that in the YB case, the affine quantum group is naturally in homogeneous grade, while in the anisotropic XXZ case it is in principal grade.}
\label{t1}
\end{center}
\end{table}
  
\section{Classical $\SU\boldsymbol{(2)}$ Lambda Models}\label{s3}

The lambda model associated to a sigma model have been defined in \eqref{sma}. 
The second term in \eqref{sma} vitiates the gauge symmetry and $A_\mu$ becomes an auxiliary Gaussian field. 
Correspondingly, the equations of motion of $A_\mu$ change from first class to second class constraints \cite{Sfetsos:2014lla}:
\EQ{
\CF^{-1}\partial_+\CF+\CF^{-1}A_+\CF&=\BOmega^TA_+\ ,\\
-\partial_-\CF\CF^{-1}+\CF A_-\CF^{-1}&=\BOmega A_-\ ,
}
where
\EQ{
\BOmega=\BId+k^{-1}\BTheta\ .
}
After integrating out the auxiliary field $A_\mu$, we can write the resulting theory as 
\be \label{eq:LambdaAction}
S_{k,{\bf \Lambda}}[{\cal F}]  = k\,S_\text{WZW}[{\cal F}] + \frac{k}{2\pi} \int d^2 \sigma \Tr\left( {\cal F}^{-1} \partial_+ {\cal F} (\BOmega  - {\textrm{Ad}}_{\cal F})^{-1} \partial_- {\cal F} {\cal F}^{-1} \right)  \ . 
\ee
This form makes it clear that as an expansion in $\BOmega^{-1}$ the theory can be interpreted as a current-current deformation of the WZW model:
\be 
S_{k,{\bf \Lambda}}[{\cal F}]  = k\,S_\text{WZW}[{\cal F}] + \frac{k}{2\pi} \int d^2 \sigma \Tr\left( {\cal F}^{-1} \partial_+ {\cal F} \BOmega^{-1}\partial_- {\cal F} {\cal F}^{-1} \right) +\cdots \ . 
\ee
The implication is that {\it if\/} the couplings flow into the UV in such a way that $\BOmega^{-1}\to0$, the lambda model  can be interpreted as a perturbed WZW CFT.

The equations of motion of the theory have a simple form when written in terms of the auxiliary field $A_\mu$:\footnote{Note that the transpose is defined  with respect to the trace: $\Tr(a\BOmega b)=\Tr(\BOmega^Ta\,b)$.}
\EQ{ \label{eq:eqmgaugefield}
\partial_+A_--\BOmega^T\partial_-A_++[\BOmega^TA_+,A_-]&=0\ ,\\
\BOmega\partial_+A_--\partial_-A_++[A_+,\BOmega A_-]&=0\ .
}

The isotropic lambda model associated to the PCM for which $\BTheta=\alpha^{-1}\BId$, gives
\EQ{
\BOmega=\lambda^{-1}\BId\ ,\qquad \lambda=\frac{k\alpha}{k\alpha+1}\ .
}
This is the model constructed and studied in \cite{Evans:1994hi}.

The XXZ version of the model has
\EQ{\label{eq:OmegaXXZ}
\BOmega=\MAT{\xi^{-1}& 0 & 0\\ 0 & \xi^{-1} & 0\\ 0 & 0 & \lambda^{-1}}\ ,
}
where 
\EQ{
\xi=\frac{k\beta}{k\beta+1}\ ,\qquad\lambda=\frac{k\alpha}{k\alpha+1}\ .
\label{zas}
}
This should be compared with the YB  version of the model for which
\EQ{\label{eq:OmegaYB}
\BOmega&=\BId+\frac1{k\alpha}(\BId-\eta\BR)^{-1}=\frac{1}{\lambda} \MAT{u_1&-u_2 & 0\\ u_2 & u_1 & 0\\ 0 & 0 &1 }\ ,\
}
where
\EQ{
u_1&=\frac{ 1+\eta^2 \lambda }{1+ \eta^2}\ ,\qquad u_2=\frac{\eta(1-\lambda )}{  1+\eta^2 }\ ,
}
and where the original sigma model couplings are
\EQ{
\lambda=\frac{k\alpha}{k\alpha+1}\ ,\qquad \frac\beta\alpha=1+\eta^2\ .
}

Now we can see that the XXZ lambda model, even in the regime $\beta>\alpha$, i.e.~$\xi>\lambda$, where the associated sigma models are equivalent up to a boundary term, is distinct from the YB lambda model. In particular, the YB lambda model breaks parity symmetry explicitly as can be seen from the fact fact that $\Theta$, entering the definition Eq.~\eqref{sma}, is not symmetric.

There is also a XYZ lambda model for which $\BOmega=\text{diag}(\lambda_i^{-1})$ with all $\lambda_i$ distinct, first constructed in  \cite{Sfetsos:2014lla}. This will be considered in more detail elsewhere \cite{forthcoming}.

\subsection{Target Spaces}\label{s3.1}

With the group element parametrized as 
\be
 {\cal F} = \MAT{  C_\phi + i S_\phi C_\psi & S_\phi S_\psi e^{- i \theta}  \\  -S_\phi S_\psi e^{i \theta} & C_\phi - i S_\phi C_\psi }  \ , 
\ee
where we have defined $S_x \equiv \sin x$ and $C_x \equiv \cos x$, the lambda theories can viewed as sigma models with target spaces of the following form
\be
\begin{aligned}
ds^2 &= \frac{k}{A_0} \left( A_1 \,d\phi^2 + A_2 \,d\psi^2 + A_3 d\theta^2 + A_4 \,d\phi \,d\psi \right) \ ,   \\
H_3 &= k \frac{A_5 }{A_0^2}  \,d\phi\wedge d\psi \wedge d\theta \ , \\
\Phi &= -\frac{1}{2} \log A_0 = - \frac{1}{2} \log \det   (\BOmega  - {\textrm{Ad}}_{\cal F} )  \  , 
\end{aligned}
\ee
where $A_i = A_i(\phi, \psi)$.
The non-trivial dilaton is due to a determinant arising from performing the Gaussian integration on the non-propagating ex-gauge fields $A_\mu$   in the path integral.
The exact functional forms are not particularly enlightening but are recorded in Appendix~\ref{a1}.  Here we note the feature, seen in other lambda deformations, that all the coordinate dependence cancels in the expression for the dilaton beta function\footnote{ To be precise $\tilde\beta^\Phi = \bar\beta^\Phi - \frac{1}{4}G^{-1}\bar\beta^G$ appears as a  coefficient  of the expectation value of the trace of the stress tensor $2\pi \langle T_a^a \rangle = \tilde\beta^\Phi  R^{(2)} + \dots$ and $\bar\beta^i$  are related to the beta-functions of couplings via a diffeomorphism generated at leading order by the derivative of the dilaton \cite{Tseytlin:1986tt,Tseytlin:1986ws,Shore:1986hk}. }, 
\be
\tilde\beta^\Phi = R+ 4 \nabla^2 \Phi - 4 (\partial \Phi)^2 - \frac{1}{12} (H_3)^2 \ . 
\ee
Explicitly we find that for the XXZ lambda model 
\EQ{ \label{eq:dilbetaxxz}
\tilde\beta^\Phi_\text{XXZ} = -\frac{2 \left(\xi ^4 (\lambda +1)-2 \xi ^2 (\lambda -1)-\lambda -1\right)}{k \left(\xi
   ^2-1\right)^2 (\lambda +1)} \ ,  
   }
in comparison to the the result obtained for the YB lambda model in  \cite{Sfetsos:2015nya} 
\EQ{\label{eq:dilbeta}
\tilde\beta^\Phi_\text{YB}=  \frac{\left(4\eta ^4-2\right) \lambda ^4-4 \left(2 \eta ^2+1\right) \lambda ^3-4
   \lambda -2}{k (\lambda -1) (\lambda +1)^3} \ .
  }
 It is noteworthy that these come out as constant despite that fact, as we will discuss later, the couplings $\xi,\eta,\lambda$ run under RG.  This is a feature of lambda models and was observed in the generalised gauged WZW models of Tseytlin \cite{Tseytlin:1993hm}. This strongly suggests that, like isotropic lambda deformations, both of these can give rise to complete solutions of type II supergravity (i.e.~define conformally invariant world sheet theories) when the theory is complemented by a similarly deformed  non-compact $\SL(2)$ WZW together with an appropriate RR sector.
 
Evidently since we have two functions of three variables one can force  $\tilde\beta^\Phi_{XXZ} $ and $\tilde\beta^\Phi_{YB} $ to be equal by relating $\eta$ and $\xi$ as 
\be
\xi^2 =\frac{(1+\eta^2) \lambda^2 }{1+ \eta^2 \lambda^2} \ . 
\ee
Later we will see this relation arising form identifying the RG invariants of the two models.
However, a more discerning comparison of $\tilde\beta^\Phi_{XXZ} $ and  $\tilde\beta^\Phi_{YB} $  can be made by recasting them in their common sigma model variables ($\alpha$ and $\beta$) making use of Eq.~\eqref{eq:abtoetat}.  The result is striking: they do {\it not\/} match!  This indicates that the XXZ and YB lambda theories are not completely equivalent. This may be surprising since the  XXZ and YB sigma models differed only by a gauge transformation of the NS two-form.  Under a conventional Buscher T-dualization, one would expect this difference to give rise to theories  related by a combination of diffeomorphism and gauge transformations after dualization. However the Sfetsos procedure  we employed  is not a dualization but instead a deformation and so there is no reason {\em a priori\/} to expect such a relationship to be the case.    The exception is in the limit $k\rightarrow \infty$, in which case the  Sfetsos procedure reduces to non-abelian T-dualization;  indeed, in this limit we find that the two expressions coincide 
\be
\tilde\beta^\Phi_\text{YB} \sim \tilde\beta^\Phi_\text{XXZ} \sim  \frac{1}{2}\beta \left(4 - \frac{\beta}{\alpha} \right)+ \mathscr O(k^{-1}) \ . 
\ee
One may recognise this as the being exactly the expected scalar curvature of the anistropic XXZ sigma model on the squashed sphere.

\subsection{Lax formalism}\label{s3.2}

Both the XXZ and YB lambda models inherit the integrability of their mother sigma models. This can be shown by constructing Lax representations of their equations of motion.

For the YB lambda model, the Lax connection was established in  \cite{Sfetsos:2015nya}. Let us first define 
\EQ{
a=\frac{1-\lambda\eta^2}{1+\lambda}
}
and functions of the spectral parameter $z$: 
\EQ{
\alpha_\pm(z) = a + \sqrt{a^2 +\eta^2} \frac{z \pm 1}{z \mp 1}\ .
}
In terms of the auxiliary gauge field $A_\mu$, the  Lax connection  equals
\be
\Lax_\pm(z) =   (\alpha(z)\pm \eta \BR  )({\bf 1} \pm \eta \BR  )^{-1}A_\pm  \ .
\ee
The sigma model limit is obtained by restoring $\lambda=k\alpha/(k\alpha+1)$ and taking $k\rightarrow \infty$ with other constants fixed.  In this limit we have  
\be
\alpha_\pm(z) \rightarrow \frac{z \pm \eta^2}{z \mp 1}  \ ,
\ee
 $A_\pm$ becomes identified with $J_\pm$ and the Lax connection reduces to that of the YB sigma model \eqref{eq:laxYB}.  Having made this connection, in order to facilitate an easier comparison to the standard form Maillet algebra, it suits us henceforward to redefine $z\rightarrow 1/z$ for the YB lambda model.

For the anisotropic XXZ lambda model, the Lax operator takes the form 
\EQ{
\Lax_\pm(z)=\sum_{a=1}^3w_a(\nu\mp z)A^a_\pm T^a\ ,
}
with
\EQ{
w_1(z)=w_2(z)=\sqrt{\frac{\lambda^2-\xi^2}{1-\lambda^2}}\cdot\frac1{\xi\sinh z}\ ,\qquad
w_3(z)=\sqrt{\frac{\lambda^2-\xi^2}{1-\xi^2}}\cdot\frac1{\lambda\tanh z} \ ,
}
and
\EQ{
\cosh^2\nu=\frac{(1-\xi)(\lambda+\xi)}{2\xi(1-\lambda)}\ .
\label{inh}
}
Note in the sigma model limit $k\to\infty$, $w_a(z)$ and $\nu$ reduce to their XXZ sigma model equivalents \eqref{h5e} and $A_\pm$ becomes identified with $J_\pm$.

\subsection{Poisson structure and symmetries}\label{s3.3}

The Poisson brackets of the lambda models are inherited from the underlying WZW model where the Kac-Moody (KM) currents are
\EQ{
\JJ_+&=-\frac k{2\pi}\big(\CF ^{-1}\partial_+\CF +\CF ^{-1}A_+\CF -A_-\big)\ ,\\
\JJ_-&=\frac k{2\pi}\big(\partial_-\CF \CF ^{-1}-\CF A_-\CF ^{-1}+A_+\big)\ ,
\label{ksq2}
}
and whose Poisson brackets take the form of two commuting classical KM algebras \cite{Bowcock:1988xr} 
\EQ{
\big\{\JJ^a_\pm(x),\JJ^b_\pm(y)\big\}&=f^{abc}\JJ_\pm^c(y)\delta(x-y)\pm\frac {k}{2\pi}\delta^{ab}\delta'(x-y)\ ,\\
\big\{\JJ^a_+(x),\JJ^b_-(y)\big\}&=0\ .
\label{km4}
}
In the present context, the $f^{abc}$ are the structure constants of the $\msu(2)$ Lie algebra.

In the YB lambda model, the spatial component of the Lax connection is written in terms of the Kac-Moody currents as \cite{Sfetsos:2015nya}
\be
\Lax(x;z) = \left( c_+(z) + d(z) \BR \right)\JJ_+(x)  + \left( c_-(z)+ d(z)  \BR \right)\JJ_-(x)\ ,
\ee
where
\EQ{
c_\pm(z) & =\mp\frac{2\pi\lambda}{k(1-\lambda^2)}(\alpha_\pm(z)-\lambda\alpha_\mp(z))\ ,\\
d(z)&=\frac{2\pi\eta\lambda}{k(1-\lambda^2)}(\lambda\alpha_+(z)+\lambda\alpha_-(z)-\lambda-1)\ .
}

The way to extract the Maillet form of the Poisson bracket of $\Lax(x;z)$ is to think of a change of variables on phase space from the KM currents $\JJ_\pm$ to the Lax operator $\Lax(z)$ and  $\Lax(w)$, for a pair of generic points $z$ and $w$. This yields precisely the form \eqref{ikk} with kernels 
\EQ{
r(z,w)&=-\frac{d(z)+d(w)}2\Big[g(z,w)\sum_{a=1}^3 T^a \otimes T^a +\sum_{a,b=1}^3 \BR _{ab} T^a \otimes T^b\Big]\ ,\\
s(z,w)&=-\frac{d(z)-d(w)}2\Big[g(z,w)\sum_{a=1}^3  T^a \otimes T^a +\sum_{a,b=1}^3 \BR _{ab} T^a \otimes T^b\Big]\ ,
}
where
\EQ{
g(z,w)=\frac{d(z)d(w)+c_\pm(z)c_\pm(w)} {c_\pm(z)d(w)- c_\pm(w)d(z)}\ ,
}
and either sign on the right-hand side can be taken.

There are two relevant limits to consider. The first is $\eta\to0$, for which the $r/s$ kernels recover the simpler form \eqref{eq:rs} with a twist function
\EQ{
\phi(z)=-\frac{k(1-\lambda^2)(1+\lambda)^2}{2\pi\lambda}\cdot\frac{1-z^2}{(1-\lambda)^2-(1+\lambda)^2z^2}\ .
\label{x7m}
}
This is the twist function quoted in \cite{Appadu:2017fff} for the isotropic lambda model.

The other interesting limit, is the sigma model limit  for which $k\to\infty$, $\lambda=k\alpha/(k\alpha+1)$, with $\alpha$ and $\eta$  fixed:
\EQ{
r(z,w)&=-\frac{\pi\alpha\eta(1+\eta^2)(z^2+w^2-2z^2w^2)}{(1-z^2)(1-w^2)}\Big[\frac{1+\eta^2 zw}{\eta(z-w)}\sum_{a=1}^3
T^a\otimes T^a+\sum_{a,b=1}^3 \BR _{ab} T^a \otimes T^b\Big]\ ,\\
s(z,w)&=-\frac{\pi\alpha\eta(1+\eta^2)(z^2-w^2)}{(1-z^2)(1-w^2)}\Big[\frac{1+\eta^2 zw}{\eta(z-w)}\sum_{a=1}^3
T^a\otimes T^a+\sum_{a,b=1}^3 \BR _{ab} T^a \otimes T^b\Big]\ ,
}
These kernels provide a different realization of the Poisson bracket algebra of the YB sigma model compared with \cite{Kawaguchi:2012gp} whose twist function we quoted in \eqref{t4r}.

For the anisotropic XXZ lambda model, the spatial component of the Lax connection is  
\EQ{
\Lax(x;z)=\sum_{a=1}^3\big(f_a(z)\JJ_+^b(x)-g_a(z)\JJ_-^a(x)\big)T^a\ ,
}
where
\EQ{
f_1(z)&=f_2(z)=\frac{2\pi}{k(1-\xi^2)}\sqrt{\frac{\lambda^2-\xi^2}{1-\lambda^2}}\big(\xi\csch(\nu+z)-\csch(\nu-z)\big)\ ,\\
f_3(z)&=\frac{2\pi}{k(1-\lambda^2)}\sqrt{\frac{\lambda^2-\xi^2}{1-\xi^2}}\big(\lambda\coth(\nu+z)-\coth(\nu-z)\big)\ ,\\
g_1(z)&=g_2(z)=\frac{2\pi}{k(1-\xi^2)}\sqrt{\frac{\lambda^2-\xi^2}{1-\lambda^2}}\big(\xi\csch(\nu-z)-\csch(\nu+z)\big)\ ,\\
g_3(z)&=\frac{2\pi}{k(1-\lambda^2)}\sqrt{\frac{\lambda^2-\xi^2}{1-\xi^2}}\big(\lambda\coth(\nu-z)-\coth(\nu+z)\big)\ ,
}
and one finds that the $r/s$ kernels are
\EQ{
r(z,w)=\frac{\phi(w)^{-1}+\phi(z)^{-1}}{\sinh(z-w)}\Pi(z,w)\ ,\qquad s(z,w)=\frac{\phi(w)^{-1}-\phi(z)^{-1}}{\sinh(z-w)}\Pi(z,w)\ ,
}
where
\EQ{
\Pi(z,w)=-T^1\otimes T^1-T^2\otimes T^2-\cosh(z-w)T^3\otimes T^3\ ,
\label{dpi}
}
and where the twist function is
\EQ{
\phi(z)=\frac{k(1+\lambda)\sqrt{1-\xi^2}}{ \pi\sqrt{\lambda^2-\xi^2}}\cdot\frac{\xi^2-\lambda+\xi(1-\lambda)\cosh(2z)}{\xi^2+\lambda-\xi(1+\lambda)\cosh(2z)}\ .
\label{y2e}
}

The isotropic limit, involves taking $\xi\to\lambda$ and $z\ll 1$ and one can verify that this gives \eqref{x7m}. The sigma model limit yields \eqref{s8i}.

\subsection{Non-local charges and infinite symmetries}

In this section, we argue that the lambda models have similar features as the sigma models where the expansion of the monodromy matrix around special points yield charges that generate infinite symmetry algebras.

Let us start our analysis with the the YB lambda model. The special points of the $r/s$ kernels are located at 
\EQ{
c_\pm(z)^2  + d(z)^2 = 0 \qquad\text{or}\qquad d(z)=0\ .
}
The former admits a pair of complex conjugate roots and the latter a pair of real roots.   We begin with the former:
\EQ{
z_*= \pm z_{I} = \pm i\frac{\sqrt{\eta ^2+1} \sqrt{\eta ^2 \lambda ^2+1}- \eta^2\lambda +1}{\eta  (\lambda +1)}\ ,
\label{cvv}
}
around which 
\EQ{
r(z,w)  &= \pm\frac{i\pi\eta\lambda}{k(1-\lambda)}\cdot\Big\{\frac{\epsilon+\tilde\epsilon}{\epsilon-\tilde\epsilon}\cdot \sum_{a=1}^3T^a\otimes T^a \mp i 
\sum_{a,b=1}^b  \BR _{ab} T^a \otimes T^b\Big\}+{\mathscr O}(\epsilon)\ ,  \\
s(z,w)&={\mathscr O}(\epsilon)\ .
}
The Lax operator itself has the expansion
\EQ{
\Lax(x;\pm z_I + \epsilon ) = \pm \frac{2 i \pi \eta \lambda }{k(1-\lambda) }  (1  \mp i \BR  )\JJ_0 (x) + O(\epsilon)\ .
}
The behaviour of the kernels here matches similar special points in the sigma model and is indicative of the existence of an affine quantum group symmetry $\mathscr U_q(\widehat{\msu(2)}_h)$ with a deformation parameter determined by the pre-factor
\EQ{
q\equiv\exp\Sigma=\exp\Big[-\frac{2\pi\eta\lambda}{k(1-\lambda)}\Big] \ .
\label{r33}
}
Later we will see that $q$ involves a renormalization group combination of the couplings. An important detail is that the affine quantum group symmetry is realized in the homogeneous gradation with the expansion parameter $\epsilon$ playing the role of the loop variable. 
 
To elucidate this structure, note that  $1  \pm i \BR $ projects onto a Borel subalgebra
\EQ{
(1\pm i\BR)T^\mp=2T^\mp\ ,\qquad (1\pm i\BR)T^\pm=0\ ,
}
so the Lax operator at the special points is 
\be
\Lax(x;\pm z_I  ) = \pm i\Sigma \left(  \JJ^3_0 (x) T^3  + 2   \JJ^\mp_0 (x)   T^\pm\right)  \ . 
\ee
Having correctly identified the special points we can continue with an analysis that replicates that of \cite{Kawaguchi:2012gp,Delduc:2013fga,Delduc:2017brb} found in the context of Yang Baxter models but novel in the context of $\lambda$ models.   In the monodromy matrix one can further factorise out the Cartan directions yielding 
\EQ{ 
T(z_I) &= \exp\left[  - \Sigma \int_{-\infty}^\infty dx\, \frak{J}^3 T^3    \right] \text{P}\overset{\longleftarrow}{\text{exp}}\Big[-  \Sigma \int_{-\infty}^\infty dx\,    \frak{J}^- T^+  \Big]\ , \\[5pt]
T(-z_I)& = \text{P}\overset{\longleftarrow}{\text{exp}}\Big[   \Sigma  \int_{-\infty}^\infty dx\,  \frak{J}^+T^-  \Big] \exp\left[    \Sigma \int_{-\infty}^\infty dx\, \frak{J}^3 T^3    \right] \ ,
 }
which further truncate due to the nilpotency of $T^\pm$.  In terms of the KM currents, we have introduced the quantities
\EQ{
\frak{J}^3(x)  &= i  \JJ^3_0 (x)  \ , \\[5pt]
\frak{J}^-(x)  &= 2i\JJ^-_0(x)   \exp\left[ + i\sqrt2 \Sigma \int^x_{-\infty}dy\,\frak{J}^3(y)\right]  \ , \\[5pt]
\frak{J}^+(x)  &= 2i\JJ^+_0(x) \exp\left[  - i \sqrt2\Sigma \int^{\infty}_{x}dy\,\frak{J}^3(y)\right] \ .
}

The Poisson brackets for these currents follow from Eq. \eqref{km4}; importantly all non ultra-local terms cancel (i.e. the brackets don't involve $\delta^\prime(x-y)$) and in particular 
\EQ{
\big\{\frak{J}^-(x)  , \frak{J}^+(y)\big\} = -\frac{2i  }{ \Sigma} \delta(x-y) \partial_x \exp  \left[- \sqrt2 i \Sigma \left( \int^{\infty}_{x} - \int^{x}_{-\infty}  \right) \frak{J}^3(y)dy \right] \ . 
} 
To see this one notices that "cross terms'' involving a single exponential in this Poisson bracket  cancel by virtue of
\EQ{
&\big\{ \JJ_0^\pm(x) , \exp\left[ i \kappa \int_\a^\b  \JJ^3_0(y) dy \right]  \big\} \\[5pt]
 =&  \pm \sqrt{2} \kappa \left( \theta(x -\alpha) - \theta(x-\beta) \right)     \JJ_0^\pm(x)  \exp\left[ i \kappa \int_\a^\b  \JJ^3_0(y) dy \right]  \ .  }
The integrals of these densities define charges  
\be
\frak{Q}^i = a^i  \int_{-\infty}^\infty dx \, \frak{J}^i(x)   \ , \quad  i \in\{3,+,-\} \ ,
\ee
which obey  
\EQ{
\{\frak{Q}^3 , \frak{Q}^\pm \}  &= \mp  a^3  \frak{Q}^\pm\ , \\ 
\{\frak{Q}^-, \frak{Q}^+ \} &= \frac{2 i  a^+ a^-}{ \Sigma}  \left( q^{  i \sqrt{2} \frak Q^3  / a^3 } - q^{-   i \sqrt{2} \frak Q^3 / a^3  }  \right) \ , 
}
and with normalisation $a^3 = i \sqrt{2} $ and $a^+=a^- $ and $(a^{+})^{-2} = 4 \Sigma^{-1} \sinh(\Sigma)$ we recover the expected QG relations.    Charges associated with the affine extension are first encountered at the next oder of the expansion of the monodromy matrix about $\pm z_I$  in accordance with the gradation. The structure of charges is presciely the same as in the sigma illustrated in Fig.~\ref{f6}.

Conservation of these charges is of course guaranteed by construction of the monodromy matrix. An explicit check is possible by making use of Eq.~\eqref{ksq2} to recast in terms of the gauge fields and the equations of motion  Eq.~\eqref{eq:eqmgaugefield}.  For instance
\EQ{
\partial_0 \frak{J}^3 & = \frac{  \eta  }{\Sigma} \left(  2\partial_+ A_-^3+2\partial_- A_+^3  + \partial_1 A_1^3 \right)\\ & =  \frac{  \eta  }{\Sigma}  \partial_1 A_1^3   =  \frac{1-\lambda}{1+\lambda} \partial_1 \JJ^3_1\ .
}
Hence $\int_{-\infty}^\infty \frak{J}^3   dx $ is a conserved quantity assuming appropriate boundary conditions.   In a similar fashion
\be
\partial_0 \frak{J}^+ = \partial_1 \bigg\{\left( \frac{1-\lambda}{1+\lambda} \JJ^+_1 + \frac{2 i \eta \lambda}{1+\lambda}    \JJ^+_0  \right)  \exp\left[ - \sqrt{2}  i \Sigma \int^{\infty}_{x}\frak{J}^3(y)dy \right]  \bigg\}  \ . 
\ee
 
In the isotropic limit, $\eta\to0$, the kernel $r$ becomes ${\mathscr O}(\epsilon)$ and the special points diverge to infinity. This kind of behaviour corresponds to the Yangian symmetry of the isotropic model which includes the gbal $\SU(2)$ symmetry of the isotropic model.

The pair of real special points are at
\EQ{
z_*=\pm z_R = \pm \frac{\lambda  \big(-2 \sqrt{\eta ^2+1} \sqrt{\eta ^2 \lambda ^2+1}+2 \eta ^2 \lambda
   +\lambda \big)+1}{1- \lambda ^2}\ ,
}
around which the leading term in the $r,s$ kernels are both ${\mathscr O}(\epsilon^0)$ and are independent of the $\eta$ parameter:
\EQ{
r(z,w)&=\pm\frac{\pi}k\cdot\frac{\epsilon+\tilde\epsilon}{\epsilon-\tilde\epsilon}\cdot\sum_{a=1}^3T^a\otimes T^a+{\mathscr O}(\epsilon)\ ,\\ s(z,w)&=-\frac\pi k\cdot\sum_{a=1}^3T^a\otimes T^a+{\mathscr O}(\epsilon)\ ,
\label{pep}
}
while the Lax operator has an expansion whose leading term is proportional to the KM currents:
 \EQ{
\Lax(x;\pm z_R + \epsilon )&=  \mp  \frac{2\pi}{k} \JJ_\pm (x) + \mathscr O(\epsilon) 
}
and so the Poisson bracket algebra becomes one of the KM algebras. The behaviour of the kernels in the neighbourhood of the special points is similar to that in \eqref{pe6} that are associated to an affine quantum symmetry. Based on this, one is tempted to identify the  deformation parameter as
\EQ{
q=\exp\Big[-\frac{i\pi}k\Big]\ .
\label{x8u}
}
This, however, this reveals an important difference. In the lambda models $k$ is effectively $\hbar^{-1}$ and so the classical limit corresponds to  $k\to\infty$. Therefore, the above deformation parameter $q\to1$ in the  classical theory. If we compare with \eqref{r33}, we see that in that case there is a consistent limit where $k\to\infty$---the classical limit---but with $q$  fixed. On top of this, if one tries to expand the monodromy matrix around these especial points one finds that the non ultra-local derivative of the delta function $\delta'(x-y)$ becomes an insurmountable problem and the expansion is not well defined. Taken together this is an indication that the symmetry structure associated to the real special points can only be understood at the quantum level. In fact, one might expect that if we first consistently quantize the theory and then take the limit $k\to\infty$, we should find a Yangian symmetry $\mathscr Y(\msu(2))$ (as described in a related context by Bernard \cite{Bernard:1990jw}). Experience with the quantum isotropic lambda model \cite{Appadu:2017fff} shows that there is indeed a quantum group symmetry with deformation parameter \eqref{x8u} but with a quantum shift by the quadratic Casimir in the adjoint (dual Coxeter number): $k\to k+2$. The deformation parameter is then a root of unity $q^{2(k+2)}=1$ and has a special representation theory that is reflected at the S-matrix level by a hidden kink structure \cite{Evans:1994hi,Appadu:2017fff}.

In the sigma model limit, the complex roots \eqref{cvv} become $z_*=\pm i/\eta$ and the affine quantum group symmetry becomes identified with that in the sigma model described in section \ref{s2}. In particular, the deformation parameter \eqref{r33} reduces to \eqref{qcl}. On the other hand, the real roots become $z_*=0$ in the limit, and $r,s$ are now ${\mathscr O}(\epsilon)$:
\EQ{
r(z,w)&=-\pi i\alpha(1+\eta^2) \cdot\frac{\epsilon^2+\tilde\epsilon^2}{\epsilon-\tilde\epsilon}\cdot\sum_{a=1}^3T^a\otimes T^a+{\mathscr O}(\epsilon^2)\ ,\\
s(z,w)&=-\pi\alpha(1+\eta^2)\cdot(\epsilon-\tilde\epsilon)\cdot\sum_{a=1}^3T^a\otimes T^a+{\mathscr O}(\epsilon^2)\ .
}
This indicates that the quantum group symmetry becomes a Yangian symmetry: indeed, the deformation parameter \eqref{x8u} goes to 1. This corresponds to the emergence of the $\SU(2)_L$ global symmetry of the sigma model.

Now we turn to the XXZ lambda model. As in the sigma model, we first change to a multiplicative spectral parameter $z\to\log z$. There are two kinds of special points: poles of the twist function \eqref{y2e} and at $z=0,\infty$.   The twist function \eqref{y2e} (after substituting $z\to\log z$) has poles at
\EQ{
\frac12(z^2+z^{-2})=\frac{\xi^2+\lambda}{\xi(1+\lambda)}\ .
}
Expanding around these special points, we find
\EQ{
r(z,w)&=\pm\frac{\pi}k\cdot\frac{\epsilon+\tilde\epsilon}{\epsilon-\tilde\epsilon}\cdot\sum_{a=1}^3T^a\otimes T^a+{\mathscr O}(\epsilon)\ ,\\ s(z,w)&=-\frac\pi k\cdot\sum_{a=1}^3T^a\otimes T^a+{\mathscr O}(\epsilon)\ ,
}
precisely as in \eqref{pep} in the YB model. So we expect the XXZ model to also to have the same affine quantum group symmetry with deformation parameter \eqref{x8u}.

The other special points are at $z_*=0,\infty$. Expanding around these points using $z=\epsilon$ and $z=\epsilon^{-1}$, respectively, gives
\EQ{
r(z,w)&=\pm\frac{2\pi\sqrt{\lambda^2-\xi^2}}{k(1-\lambda)\sqrt{1-\xi^2}}\cdot\frac{\epsilon\tilde\epsilon}{\epsilon^2-\tilde\epsilon^2}\sum_{a=1}^3T^a\otimes T^a+{\mathscr O}(\epsilon)\ ,\\
s(z,w)&={\mathscr O}(\epsilon^2)\ .
}
This is identical to behaviour of the kernels in the XXZ sigma model \eqref{p9p} and is therefore associated with an affine quantum group symmetry with
\EQ{
q=\exp\Big[-\frac{2\pi\sqrt{\xi^2-\lambda^2}}{k(1-\lambda)\sqrt{1-\xi^2}}\Big]\ .
\label{u7a}
}
Here one finds in an expansion of the Lax around $z=0,\infty$ that the term entering at order ${\mathscr O}(\epsilon^0)$ contains the current $ \JJ^3_0$ whilst the remaining currents are found at  ${\mathscr O}(\epsilon^{\pm1})$ .

So it is clear that the XXZ lambda model and YB lambda model have the same symmetries, i.e.~a pair of affine quantum (loop) groups. However, the loop group is realized in a different gradation. For the YB case, we have the untwisted affinization which corresponds to the homogeneous gradation $\mathscr U_q(\widehat{\msu(2)}_h)$, while in the XXZ case, it is  the principal gradation $\mathscr U_q(\widehat{\msu(2)}_p)$. This difference will prove crucial for the S-matrices to be discussed in section \ref{s5}. The symmetries are summarized in Table \ref{t2}. 
\begin{table}
\begin{center}
\begin{tikzpicture}[scale=1]
\node at (-4.5,0) {\phantom{.}};
\node at (0,0) {$\lambda$-isotropic};
\node at (0,1) {$\lambda$-YB};
\node at (0,-1) {$\lambda$-XXZ};
\node at (3,0) {$\mathscr U_{q}(\widehat{\msu(2)})$};
\node at (3,1) {$\mathscr U_{q}(\widehat{\msu(2)})$};
\node at (3,-1) {$\mathscr U_{q}(\widehat{\msu(2)})$};
\node at (6,0) {$\mathscr Y(\msu(2))$};
\node at (6,1) {$\mathscr U_q(\widehat{\msu(2)}_h)$};
\node at (6,-1) {$\mathscr U_q(\widehat{\msu(2)}_p)$};
\node (a1) at (10,-0.8) {\footnotesize $q=\exp\big[-\frac{2\pi\sqrt{\xi^2-\lambda^2}}{k(1-\lambda)\sqrt{1-\xi^2}}\big]$};
\node (a3) at (9.5,0.8) {\footnotesize $q=\exp\big[-\frac{2\pi\eta\lambda}{k(1-\lambda)}\big]$};
\draw[->] (a3) -- (7.3,1);
\draw[->] (a1) -- (7.3,-1);
\draw[very thick] (-1,-1.5) -- (7.5,-1.5);
\draw[very thick] (-1,2.5) -- (7.5,2.5);
\draw[thick] (-1,1.5) -- (7.5,1.5);
\node at (0,2) {Model};
\node at (3,2) {Left symm.};
\node at (6,2) {Right symm.};
\node (a2) at (1,-2) {\footnotesize $q=\exp\big[-\frac{i\pi}k\big]$};
\draw[->] (a2) -- (2.3,0.7);
\draw[->] (a2) -- (2.3,-0.3);
\draw[->] (a2) -- (2.3,-1.3);
\end{tikzpicture}
\caption{\footnotesize The symmetries of the lambda models. Here, ``left" and ``right" refer to the relation with the sigma model symmetries in the sigma model limit $k\to\infty$.}
\label{t2}
\end{center}
\end{table}
   
\section{Renormalization Group Flow}\label{s4}
  
Ultimately the question of whether the lambda models define consistent continuum QFTs relies on the existence of an RG fixed point in the UV. We can investigate this question in perturbation theory by calculating the one-loop beta function of the theories. We shall find that the YB and anisotropic XXZ theories in the regime $\beta>\alpha$, precisely the models with affine quantum group symmetry with a real $q$, have an exotic cyclic type RG behaviour. The interpretation of this is quite subtle but has been considered in the closely related context of perturbed WZW models in \cite{Leclair:2003xj,LeClair:2003hj,LeClair:2004ps}.

\subsection{The sigma model RG flow}\label{s4.1}
 
The RG flow is not affected by the boundary terms in the YB case and so are the same in both the XXZ and YB cases. Writing the flows on the couplings $\alpha$ and $\beta$, we have at the one-loop level
\EQ{
\mu\frac{d\beta}{d\mu}=\beta^2\Big(\frac{\beta}{\alpha}-2\Big)\ ,\qquad \mu\frac{d\alpha}{d\mu}=-\beta^2\ ,
\label{smrg}
}    
where $\mu$ is the RG running energy scale.

These flow equations have an invariant
\EQ{
\gamma^{\prime2}=\frac{1}{4\alpha(\alpha-\beta)}\ ,
\label{rgp}
}
which is real for the XXZ model with $\beta<\alpha$ and purely imaginary in the YB and the XXZ models with $\beta>\alpha'$. In the latter case, we will write $\gamma'=i\sigma$ with $\sigma$ real. The RG flows will depend crucially on whether $\gamma'$ is real or $\sigma$ is real.

A key observation now presents itself: the deformation parameter of the affine quantum group symmetry of the sigma model \eqref{qcl} is an RG invariant! In fact,
\be
q = \exp\big[- i\pi/\gamma^\prime\big] =\exp\big[-\pi/\sigma\big]\ . 
\label{y6t}
\ee
This is very striking and explains why the quantum group symmetry can survive the classical limit and become manifest at the Poisson bracket level. There is an important difference between the XXZ model in the regime $\beta>\alpha$ and the YB model compared with the XXZ model in the regime $\alpha>\beta$. In the former $q$ is real while in the latter $q$ is a complex phase. This distinction will correlate with entirely different RG behaviour.

Turning to the RG flows, for $\gamma'\in\mathbb R$, we can solve the beta function equations to find
\EQ{
\frac1\alpha=2\gamma'\tanh \zeta\ ,\qquad\frac1\beta=\gamma'\sinh (2\zeta)\ ,
\label{pp1}
}
where $\zeta$ is related to the RG scale implicitly via
\EQ{
\zeta+\frac12\sinh(2\zeta)=\frac1{\gamma'}\log\frac\mu\Lambda\ .
\label{pp2}
}
This shows that the theory has a good UV limit $\mu\to\infty$ for which $\beta\to0$ and $\alpha\to(2\gamma')^{-1}$, a constant.
The theories in this regime lie in the class of Fateev's SS models \cite{Fateev:1996ea} and are conjectured to have an S-matrix that has the product form \eqref{w8e}. 

On the other hand, in the XXZ model, in the regime $\beta>\alpha$, and the YB model, the RG flow does not have a UV safe limit. In order to see this, note that the solution  \eqref{pp1} and \eqref{pp2} can be analytically continued to cover this regime:
\EQ{
\frac1\alpha=2\sigma\tan \zeta\ ,\qquad\frac1\beta=\sigma\sin(2\zeta)\ ,
\label{pp3}
}
where
\EQ{
\zeta+\frac12\sin(2\zeta)=\frac1\sigma\log\frac\mu\Lambda\ .
\label{pp4}
}
In this case, the RG flow appears to follow a cycle. However, the cycle passes outside the perturbative regime (small $\alpha$ and $\beta$) and so it is not clear that the one-loop result can be trusted. Theories with RG limit cycles have been the subject of a lot of interest and there are several physical applications (see the review \cite{Bulycheva:2014twa} and references therein). 

The relation between the reality of $q$ and RG behaviour seems to be quite general:
\begin{center}
\begin{tikzpicture}[scale=1]
\draw[-] (-3.5,0.5) -- (3.5,0.5) -- (3.5,-1.5) -- (-3.5,-1.5) -- (-3.5,0.5);
\node at (0,0) {$|q|=1\qquad\implies~~~\quad \text{UV safe}$};
\node at (0,-1) {$q\in\mathbb R\qquad~~\implies\quad~ \text{cyclic RG}$};
\end{tikzpicture}
\end{center}

\subsection{Yang Baxter lambda model RG flow}\label{s4.2}

In the YB lambda model, the one loop beta functions are
\cite{Sfetsos:2015nya}
\EQ{\label{eq:YBflow}
\mu\frac{d\lambda}{d\mu}&=-\frac4k\cdot\frac{\lambda^2(\eta^2+1)(\eta^2\lambda^2+1)}{(\lambda+1)^2}\ ,\\
\mu\frac{d\eta}{d\mu}&=-\frac4k\cdot \frac{  \eta  \lambda \left(\eta^2+1\right) \left(\eta ^2 \lambda^2+1\right)}{  (\lambda -1)  (\lambda +1)^2}\ .
}
It is important to note that in the lambda model, the loop counting parameter is the inverse WZW model level $k^{-1}$ and the beta function is exact at this order in $k^{-1}$ as a function of the couplings $\lambda$ and $\eta$. One can readily verify that there is an RG invariant combination
\EQ{\label{eq:invt}
\sigma=\frac{k(1-\lambda)}{2\eta\lambda}\ .
}
Once again we see that the quantum deformation parameter that we established at the classical level \eqref{r33} is an RG invariant.

In the YB lambda model we note an important duality symmetry on couplings and fields 
\be
\lambda \to \frac{1}{\lambda}  \ , \qquad \eta \to \eta \lambda  \ ,  \qquad k\to - k \ , \qquad  {\cal F}\to{\cal F}^{-1} \ .
\label{awq}
\ee
This leaves the worldsheet action Eq.~\eqref{eq:LambdaAction} classically invariant and extends the similar duality symmetry seen in the isotropic case.  Under this transformation we note that the target space metric and two form are necessarily invariant however the dilaton receives a constant shift:
\be
\Phi\to  \Phi  -\frac{1}{2} \log\Big[ - \frac{(1+\eta^2) \lambda^3 }{1+\eta^2 \lambda^2} \Big]\ . 
\ee
The dilaton beta function Eq.~\eqref{eq:dilbeta} is invariant under this mapping.  
%

\pgfdeclareimage[interpolate=true,width=12cm]{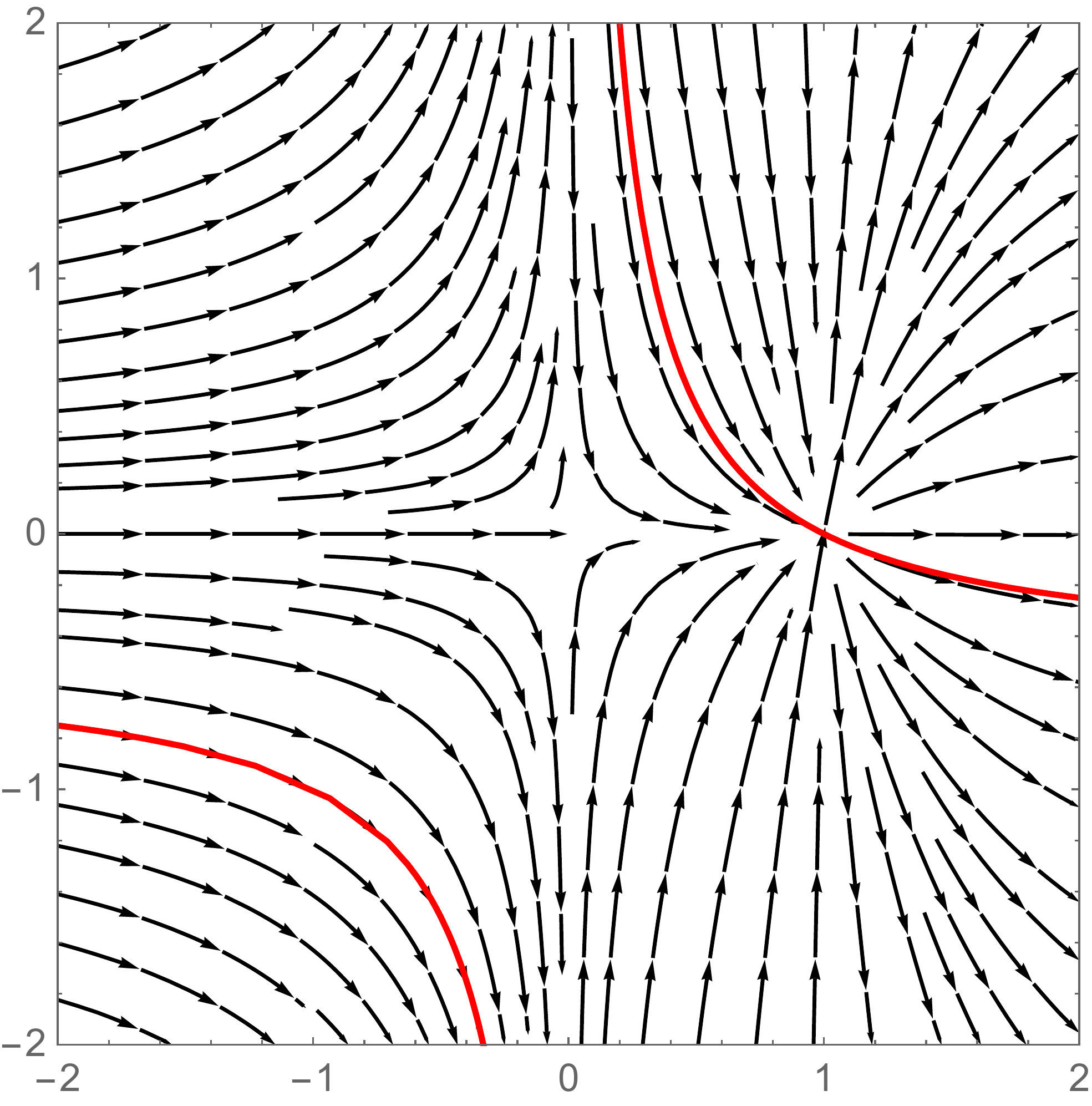}{fs1}
\begin{figure}
\begin{center}
\begin{tikzpicture}[scale=0.75]
\pgftext[at=\pgfpoint{0cm}{0cm},left,base]{\pgfuseimage{fs1}} 
\node at (6.29,-0.4) {$\lambda$};
\node at (-0.7,6.29) {$\eta$};
\filldraw[blue] (6.29,6.29) circle (4pt);
\end{tikzpicture}
\caption{\footnotesize The RG flow of the YB lambda model (flows towards the IR). The WZW fixed point is the blue dot in the middle. The red curved is an example of a cyclic trajectory which has a jump from $\eta=+\infty$ to $-\infty$ at $\lambda=0$ and a jump from $\lambda=-\infty$ to $\lambda=+\infty$.}
\label{f2}
\end{center}
\end{figure}
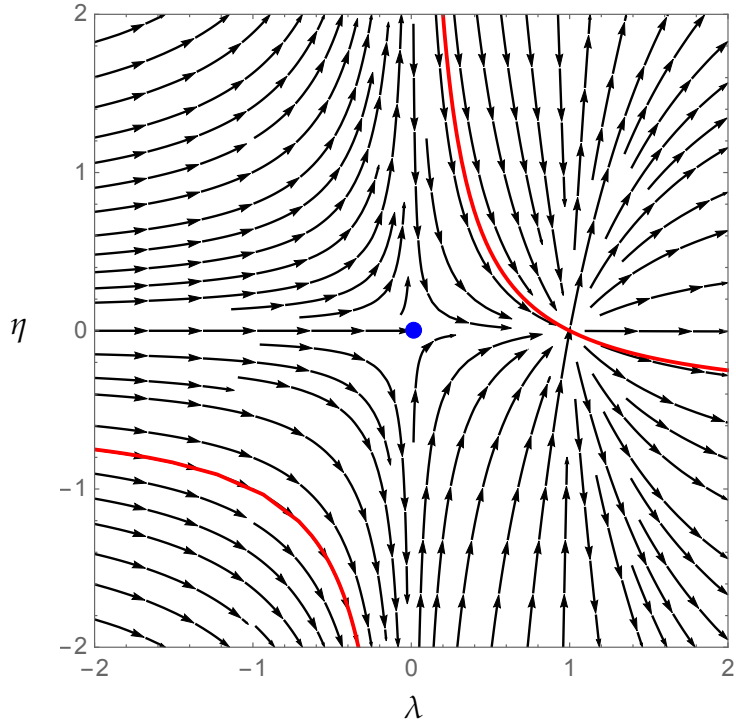

We can use the RG invariant to eliminate $\eta$ to get a single equation for $\lambda$:
\EQ{
\mu\frac{d\lambda}{d\mu}=-\frac1{4k}\cdot\frac{(k^2(1-\lambda)^2+4\sigma^2)(k^2(1-\lambda)^2+4\sigma^2\lambda^2)}{\sigma^4(1+\lambda)^2}\ .
\label{zmm}
}
Integrating gives $\lambda$ implicitly in terms of the RG scale $\mu$
\EQ{
&-2\sigma\Big(\tan^{-1}\frac{k(\lambda-1)}{2\sigma}+\tan^{-1}\frac{k^2(\lambda-1)+4\sigma^2\lambda}{2\sigma k}\Big)\\ &\qquad\qquad\qquad\qquad+k\log\frac{k^2(\lambda-1)^2+4\sigma^2}{k^2(\lambda-1)^2+4\sigma^2\lambda^2}=4\log\frac\mu\Lambda\ .
\label{wep}
}

The RG flows are shown in Fig.~\ref{f2}. Apart from discontinuities at infinity, the flow follows a cycle. The jumps are seen to be continuous in terms of the dual couplings in \eqref{awq} and so we interpret the flows as following a physically continuous set of theories. In addition, the beta function \eqref{zll} has a pole at $\lambda=1$ but the flow is perfectly well defined through it.
For a cyclic RG flow, a key quantity is change in the energy scale $\mu$ as the flow goes around one cycle \cite{Leclair:2003xj,LeClair:2003hj,LeClair:2004ps}. 
This follows easily from \eqref{wep}: around a cycle each of the arctan functions jump by $\pi$ and so around a complete cycle  the energy scale changes by a factor
\EQ{
\mu\longrightarrow \mu\exp[\pi\sigma]\ .
}

Given the famous {\em c-theorem} of Zamolodchikov \cite{Zamolodchikov:1986gt}, the presence of RG cycles may come as some surprise since na\"\i vely these seem to forbid the existence of a monotonic function along the flow.  To assuage anxiety we note the couplings as functions of scale are multi-sheeted can this can allow for a (unbounded) monotonic function that jumps sheets as a cycle is traversed (see \cite{Curtright:2011qg} for a toy model exhibiting this fact).    One may further wonder about the robustness of these cycles as the one-loop RG equations are employed in domains where the couplings are not small; however one should keep in mind that the loop counting parameter $k^{-1}$ does remain small.   Nonetheless, further study is required to definitively conclude the existence of such behaviour; it may be that the theory in this domain should be viewed only as an effective theory with a cut-off that is necessarily encountered before an RG cycle can be completed.  We will return to this in the next section and comment further about this possibility in the conclusion.  

\subsection{Anisotropic XXZ lambda model RG flow}\label{s4.3}

Now we analysis RG flow in the XXZ lambda model.
The RG flow of the two couplings follows from the general formula in \cite{Sfetsos:2014jfa}:
\EQ{
\mu\frac{d\xi}{d\mu}&=\frac4k\cdot\frac{\xi(\xi^2-\lambda)}{(1-\xi^2)(\lambda+1)}\ ,\\ \mu\frac{d\lambda}{d\mu}&=-\frac 4k\cdot\frac{\xi^2(1-\lambda)^2}{(1-\xi^2)^2}\ .
\label{rg2}
}  
Note in the sigma model limit, $k\to\infty$ we get precisely the sigma model RG flow \eqref{smrg} when we use \eqref{zas}.

The RG flow in this case also has an invariant
\EQ{
\gamma^{\prime2}=\frac{k^2}4\cdot\frac{(1-\xi^2)(1-\lambda)^2}{\lambda^2-\xi^2}\ .
\label{rgi}
}
We have used the same notation $\gamma'$ for the RG invariant here because in the sigma model limit, \eqref{zas} with $k\to\infty$, we have 
\EQ{
\gamma^{\prime2}\longrightarrow \frac1{4\alpha(\alpha-\beta)}\ ,
\label{fr3}
}
precisely, as it must be, the RG invariant of the sigma model \eqref{rgp}.

The XXZ lambda model also has a duality symmetry that takes
\be
\lambda \to \frac{1}{\lambda}  \ , \qquad\xi \to \frac1\xi  \ ,  \qquad k \to - k \ , \qquad  {\cal F}\to {\cal F}^{-1} \ .
\ee
The RG invariant is also invariant under this symmetry. These kinds of duality symmetries have previously been investigated in the context of current-current deformations of WZW models in \cite{Gerganov:2000mt,Bernard:2001sc}.

\pgfdeclareimage[interpolate=true,width=14cm]{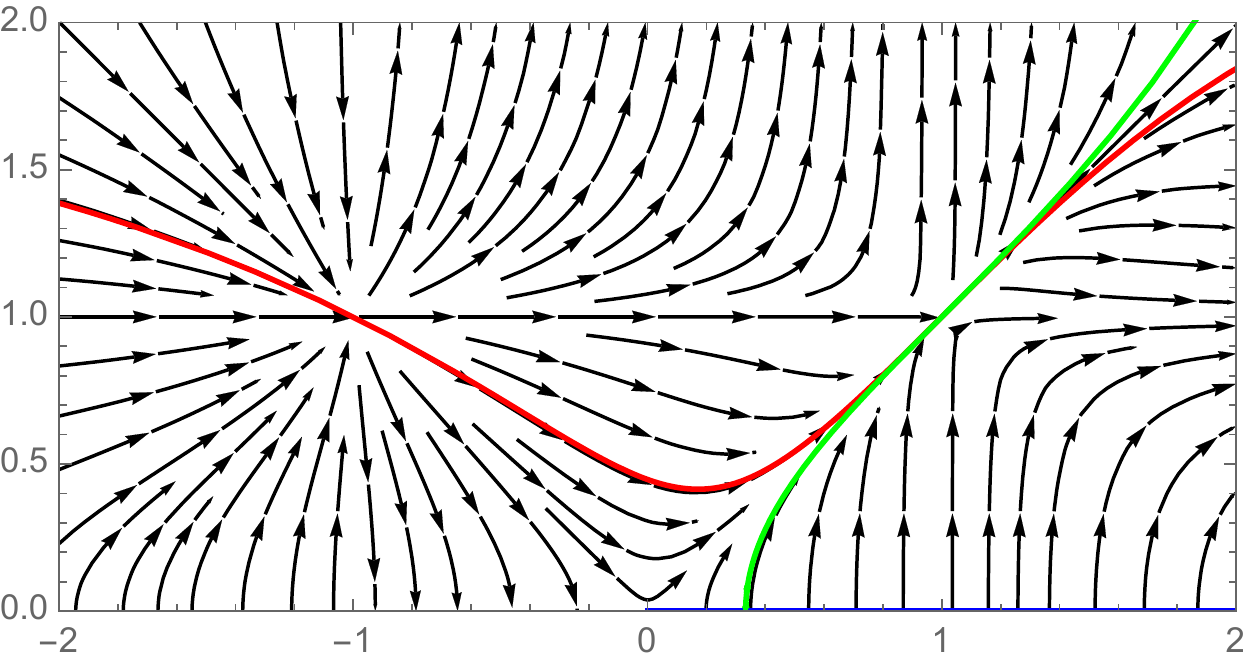}{fs2v2}
\begin{figure}
\begin{center}
\begin{tikzpicture}[scale=1]
\pgftext[at=\pgfpoint{0cm}{0cm},left,base]{\pgfuseimage{fs2v2}} 
\node at (-0.4,4) {$\xi$};
\node at (7.35,-0.4) {$\lambda$};
\node at (8.6,0.2) {$\lambda_*$};
\filldraw[blue] (7.35,0.75) circle (4pt);
\end{tikzpicture}
\caption{\footnotesize The RG flow (to the IR) of the XXZ lambda model. The WZW fixed point is identified by the blue blob. The blue line is a line of UV fixed points. The green curve is a UV safe trajectory that has $\gamma'\in\mathbb R$. The red curve is a cyclic RG trajectory with $\gamma'=i\sigma$, $\sigma\in\mathbb R$. The trajectory has a jump in the coupling $\lambda$ from $-\infty$ to $\infty$, but is continuous in the dual coupling $1/\lambda$.}
\label{f3}
\end{center}
\end{figure}
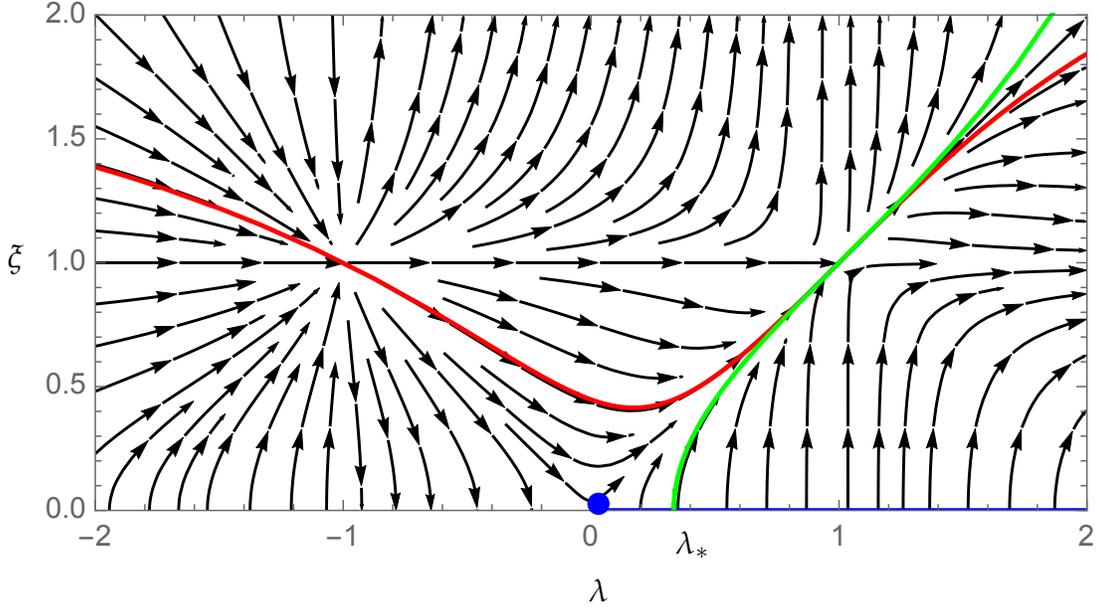

There are two distinct types of RG flow that depend on whether $\gamma'$ is real or imaginary which are ``UV safe" and "cyclic", respectively. The RG flows are shown in Fig.~\ref{f3}. We can use the RG invariant to solve for $\xi$ and substituting into \eqref{rg2}, we can write a single equation for $\lambda$,
\EQ{
\mu\frac{d\lambda}{d\mu}=-\frac1{4k}\cdot\frac{(k^2(1-\lambda)^2-4\gamma^{\prime2})(k^2(1-\lambda)^2-4\gamma^{\prime2}\lambda^2)}{\gamma^{\prime4}(1+\lambda)^2}\ .
\label{zll}
}
We will soon exploit the fact that this is identical with the RG flow equation for the YB lambda model \eqref{zmm} with $\gamma'\to i\sigma$. 

 Since $\tilde{\beta}^\Phi$ \cite{Tseytlin:1987bz,Tseytlin:2006ak} can be thought of as a generalised central charge function (and its integral $S_{eff} = \int \sqrt{G} e^{-2\Phi}  \tilde{\beta}^\Phi$ the central charge action) is natural to study its property along the RG flows.  In principle one simply needs to substitute the solution of the RG equations into the expression \eqref{eq:dilbetaxxz}. In practice given the implicit form for the solutions  to eq.~\eqref{rg2}  it is expedient to proceed  numerically and study the evolution along   for instance   the green and red trajectories of Fig.~\ref{f3}. On the UV safe trajectory  one finds $\tilde{\beta}^\phi$  decreases monotonically except at one point (the saddle point in Fig.~\ref{f3} where $\lambda=1$, $\xi =1$ ) where $\tilde{\beta}^\phi$ jumps from $-\infty$  to $+\infty$.   Similarly on the UV cyclic red trajectory $\tilde{\beta}^\Phi$  decreases monotonically except at two points (  where $\lambda=\pm 1$ and  $\xi =1$ ).  Being a function of cyclic functions of RG time in this case $\tilde{\beta}^\Phi$ returns to itself after a complete cycle.    Thus with the exception of isolated points in which  $\tilde{\beta}^\Phi$ is discontinuous, it is elsewhere monotonic.  Although these points look rather innocuous  in the RG flow--they are saddles in the $\xi,\lambda$ plane--they are distinguished from the sigma model perspective as locations in which the determinant of the target space metric changes sign.

The UV safe regime, corresponds to $\gamma'\in\mathbb R$, so quantum group parameter $q$ a complex phase. In this region, as  the flow runs backwards towards the UV, $\xi$ goes to zero while $\lambda$ goes to a constant that we denote $\lambda_*$ which is determined by the RG invariant via
\EQ{
\lambda_*=\frac k{k+2\gamma'}\ .
}
These flows have a safe UV limit and in the UV, we can expand the couplings in powers of $q=(\Lambda/\mu)^\nu$, where
\EQ{
\nu=\frac{4\lambda_*}{k(1+\lambda_*)}=\frac2{\gamma'+k}\ ,
\label{nen}
}
and $\Lambda$---the lambda parameter---is the dynamically generated mass scale. The series are of the form
\EQ{
\lambda=\lambda_*+\sum_{n=1}^\infty \lambda_nq^{2n}\ ,\qquad \xi=\sum_{n=1}^\infty\xi_nq^{2n-1}\ .
}

The points $\lambda=\lambda_*$ varying and $\xi=0$ parametrize a line of UV fixed points shown in blue in Fig.~\ref{f3}. For small couplings the action takes the form of a current-current perturbation of the WZW model,\footnote{In the WZW model, the currents are $\JJ_+=-k/(2\pi)\CF^{-1}\partial_+\CF$ and $\JJ_-=k/(2\pi)\partial_-\CF\CF^{-1}$.}
\EQ{
k\,S_\text{WZW}[\CF]-\frac{2\pi}{k}\int d^2x\,\Big(\xi\JJ_+^1\JJ_-^1+\xi\JJ_+^2\JJ_-^2+\lambda\JJ_+^3\JJ_-^3\big)\ .
}
The fixed line corresponds to just turning on the $\JJ_+^3\JJ_-^3$ perturbation. 

It is known that the $\SU(2)$ WZW model does lie on a line of fixed points. In order to see this, one uses the fact that the $\SU(2)$ WZW model at level $k$ can be realized as a compact scalar on a circle of radius $R$ coupled to $\mathbb Z_k$ parafermions \cite{Fateev:1985mm,Bernard:1990ti}. The WZW point has the critical radius
\EQ{
R_*=\sqrt{\frac1{2k}}\ .
}
The scalar field determines the one of the components of the currents via
\EQ{
\JJ_\pm^3=2i\sqrt{k}\partial_\pm\varphi\ .
}
The critical line emerges because the model remains critical as we change 
the radius. This corresponds to adding the term $\JJ_+^3\JJ_-^3$ to the action which is clearly equivalent to the $\lambda$ coupling at $\xi=0$ for small $\lambda$.
Adding $\JJ_+^1\JJ_-^1+\JJ_+^2\JJ_-^2$ on top of this, gives an integrable massive deformation corresponding to turning on $\xi$ (for $\lambda_*>0$).

So in the UV limit, $\xi\to0$ and $\lambda$ goes to a constant $\lambda_*$ and one has 
\EQ{
\lambda_*=\frac{R_*^2-R^2}{R_*^2+R^2}\ .
\label{pol}
}

Bernard and LeClair \cite{Bernard:1990ti}  identify the S-matrix of the perturbed theory, the so-called ``fractional sine-Gordon" theory, as
\EQ{
S_{\lambda\text{-XXZ}}(\theta)=S_\text{RSOS}(\theta;k)\otimes S(\theta;\gamma')\ ,
\label{h90}
}
where the second block is the sine-Gordon soliton S-matrix and the first factor describes additional kink quantum numbers of the states. 
The sine-Gordon S-matrix with coupling is
\EQ{
\gamma'=\frac{kR^2}{R_*^2-R^2}=\frac{k(1-\lambda_*)}{2\lambda_*}\ .
\label{sgc}
}
This is exactly the RG invariant we defined in \eqref{rgi} and explains our earlier notation. 

Now we turn to the regime of imaginary $\gamma'=i\sigma$, for $\sigma\in\mathbb R$. Note that the RG equation \eqref{zll} with $\gamma'\to i\sigma$ is precisely the same as the RG equation \eqref{zmm} in the YB lambda model. This is significant and suggests that YB lambda model and XXZ in the cyclic regime are closely related.

The solution for $\lambda$ in terms of $\eta$ is he same as \eqref{wep} and there is an RG cycle.
A typical trajectory is shown in Fig.~\ref{f3} in red. Just as in the YB lambda model, the trajectory follows a closed cycle which involves a jump from $+\infty$ to $-\infty$ in $\lambda$ which is continuous in the dual coupling $1/\lambda$. 

\section{S Matrices}\label{s5}

In this section, we make informed conjectures for the S-matrices of the generalized lambda and sigma models. In order to pin down the S-matrix there are some important pieces of information to take into account:
\begin{enumerate}
\item The S-matrix of the isotropic lambda model associated to the PCM takes the form of a product of the rational, i.e.~$\mathscr Y(\msu(2))$ invariant S-matrix, and an affine quantum group $\mathscr U_q(\widehat{\msu(2)})$ RSOS kink S-matrix \cite{Evans:1994hi,Appadu:2017fff}:
\EQ{\label{eq:SXXX}
S_{\lambda\text{-PCM}}(\theta)=S_\text{RSOS}(\theta;k)\otimes S_\text{SU(2)}(\theta)\ .
}
 In the limit, $k\to\infty$ the RSOS factor becomes the rational limit of the unrestricted  $S_\text{SOS}(\theta)$ which is itself the vertex-to-IRF transform of the $\SU(2)$ invariant S-matrix block. This manifests at the S-matrix level that the $k\to\infty$ limit of the lambda model is the non-abelian T-dual of the PCM:
\EQ{
\begin{tikzpicture}[scale=1]
\node at (-4.15,2.5) {$\overbracket{S_{\text{RSOS}}(\theta;k)\otimes S_{\SU(2)}(\theta)}^{\lambda\text{-model}}$};
\draw[thick,<->] (-4,1.9) -- (-4,0.5);
\node at (-3,1.25) {$k\to\infty$};
\node at (0,0) {$S_{\text{SOS}}(\theta)\otimes S_{\SU(2)}(\theta)\overset{\text{\footnotesize NAT-duality}}{\xleftrightarrow{\hspace*{3cm}}}\underbracket{S_{\SU(2)_L}(\theta)\otimes S_{\SU(2)_R}(\theta)}_{\text{PCM S-matrix}}$};
\end{tikzpicture}
\label{x99}
}
\item The XXZ sigma model with $\gamma'\in\mathbb R$ lies in the class of SS models of Fateev \cite{Fateev:1996ea}. The S-matrix is then known to have the product form \eqref{w8e}
where $\gamma'$ is the RG invariant related to the UV limit of the coupling $\lambda$ as in \eqref{rgi}.
\item The YB lambda model breaks parity while the XXZ model preserves parity.
\item As described in sections \ref{s2} and \ref{s3}, the classical sigma and lambda models have Poisson bracket realizations of the affine quantum group $\mathscr U_q(\widehat{\msu(2)})$ where $q$ is related to the RG invariants as in \eqref{qcl}, \eqref{r33} and \eqref{u7a}.
\item For the theories with cyclic RG flow with a periodicity $\mu\to\mu e^{\pi\sigma}$, it is expected that the S-matrices at high energy have a periodicity in rapidity to match \cite{Leclair:2003xj}:
\EQ{
S(\theta+\pi\sigma)=S(\theta)\ ,\qquad(\theta\gg1)\ .
\label{ppr}
}
The intuition here is that in the UV at energy scales $E\gg m$, the RG cycle behaviour requires that the theory has a discrete scaling symmetry $E\to E\exp(\pi\sigma)$. But for a particle state with $E\gg m$, i.e.~$\theta\gg1$, we have $E\approx me^\theta/2$ and so the scaling symmetry corresponds to a rapidity shift $\theta\to\theta+\pi\sigma$.
\end{enumerate}

\subsection{Quantum group S-matrix: $\boldsymbol q$ complex phase}\label{s5.1}

Before making our S-matrix conjectures, there are some general features of S-matrix theory in the integrable context to take into account. S-matrices for relativistic integrable QFTs with degenerate particle multiplets are built out of solutions to the Yang-Baxter equation, for which quantum groups provide an algebraic framework. For present purposes, we will be interested in the quantum group deformation of the affine (loop) Lie algebra $\mathscr U_q(\widehat{\msu(2)})$. We start with the case when $q$ is a complex phase in which case the S-matrix describes the scattering of solitons in the sine-Gordon theory \cite{Zamolodchikov:1978xm}.

The S-matrix in an must satisfy some important identities (described, for example, in the lectures \cite{Dorey:1996gd}):
\begin{enumerate}
\item {\it Factorization\/}. Due to integrability, there is no particle production and the complete S-matrix is determined by the $2\to2$ body S-matrix elements, as illustrated in Fig.~\ref{f1}.
\item {\it Analyticity\/}. The S-matrix is an analytic function of the complexified rapidity with poles along the imaginary axis $0<\IM\theta<\pi$ associated to stable bound states. Since there is no particle creation in an integrable field theory there are no particle thresholds, however, there can be anomalous thresholds in the form of additional, usually higher order, poles 
\item {\it Hermitian analyticity\/}
\EQ{
S_{ij}^{kl}(\theta^*)^*=S_{kl}^{ij}(-\theta)\ .
\label{ha}
}
\item {\it Unitarity\/}
\EQ{
\sum_{kl}S_{ij}^{kl}(\theta)S_{mn}^{kl}(\theta)^*=\delta_{im}\delta_{jn}\ ,\qquad \theta\in\mathbb R\ .
\label{qft}
}
\item {\it Crossing\/}
\EQ{
S_{ij}^{kl}(\theta)={\cal C}_{kk'}S_{k'i}^{lj'}(i\pi-\theta){\cal C}^{-1}_{j'j}=S_{\bar ki}^{l\bar j}(i\pi-\theta)\ ,
\label{cross}
}
where ${\cal C}$ is the charge conjugation matrix.
\end{enumerate}
Unitarity is implied by Hermitian analyticity and the {\it braiding relation\/}
\EQ{
\sum_{kl}S_{ij}^{kl}(\theta)S_{kl}^{mn}(-\theta)=\delta_{im}\delta_{jn}\ ,
\label{bru}
}
which is more natural in the context of quantum groups.

In the present context, the basis states $\ket{m}$ transform in the spin $\frac12$ representation of $\msu(2)$, or the quantum group $\mathscr U_q(\msu(2))$, with $m=\pm\frac12$. The 2-body S-matrix is a map, or {\it intertwiner\/},
\EQ{
S(\theta):\qquad V(\theta_1)\otimes V(\theta_2)\longrightarrow V(\theta_2)\otimes V(\theta_1)\ ,
}
where $V(\theta)$ is the vector space spanned by the states $\ket{\pm\frac12,\theta}$.
Here, the rapidity of the states is $\theta_i$, and $\theta=\theta_1-\theta_2$ is the rapidity difference. The S-matrix takes the form
\EQ{
S(\theta)=f(\theta)\check R(x(\theta))\ ,
\label{lala}
}
where $x(\theta)=e^{c\theta}$, $c$ to be determined, and $\check R(x)$ is the $R$-matrix of the affine quantum group $\mathscr U_q(\widehat{\msu(2)})$
\EQ{
\check R(x)=xT^{-1}-x^{-1}T\ ,
\label{we4}
}
where, on a basis for $V$, $\Be_1\equiv\ket{\frac12}$ and $\Be_2\equiv\ket{-\frac12}$,
\EQ{
T\Be_i\otimes\Be_j=\begin{cases} q^{-1}\Be_i\otimes\Be_i & i=j\ ,\\ (q^{-1}-q)\Be_i\otimes\Be_j+\Be_j\otimes \Be_i & i>j\ ,\\ \Be_j\otimes \Be_i &i<j\ ,\end{cases}
\label{ier}
}
is a generator of the Hecke algebra (the commutant of the quantum group acting on tensor products) and obeys
\be\label{eq:Hecke}
(T+ q ) \cdot (T- q^{-1} ) = 0  \ . 
\ee

In \eqref{lala}, $f(\theta)$ is a scalar factor which is needed to ensure that the S-matrix satisfies the S-matrix constraints of crossing and unitarity. Based on matrix form of $\check R$, there are four basic processes; {\it identical particle\/}, {\it transmission\/} and and two kinds of {\it reflection\/}:
\EQ{
S_I(\theta)&=\begin{tikzpicture}[baseline=-0.65ex,scale=0.5]
\filldraw[black] (0,0) circle (4pt);
    \draw[->] (-0.6,-0.6) -- (0.6,0.6);
    \draw[<-] (-0.6,0.6) -- (0.6,-0.6);
    \node at (-1.3,1.3) (a1) {$\pm\frac12$};
       \node at (1.3,-1.3) (a2) {$\pm\frac12$};
           \node at (1.3,1.3) (a3) {$\pm\frac12$};
    \node at (-1.3,-1.3) (a4) {$\pm\frac12$};
\end{tikzpicture}=f(\theta)(xq-q^{-1}x^{-1})\ ,\\
S_T(\theta)&=\begin{tikzpicture}[baseline=-0.65ex,scale=0.5]
\filldraw[black] (0,0) circle (4pt);
    \draw[->] (-0.6,-0.6) -- (0.6,0.6);
    \draw[<-] (-0.6,0.6) -- (0.6,-0.6);
    \node at (-1.3,1.3) (a1) {$\mp\frac12$};
       \node at (1.3,-1.3) (a2) {$\mp\frac12$};
           \node at (1.3,1.3) (a3) {$\pm\frac12$};
    \node at (-1.3,-1.3) (a4) {$\pm\frac12$};
\end{tikzpicture}=f(\theta)(x-x^{-1})\ ,\\
S_R^\pm(\theta)&=\begin{tikzpicture}[baseline=-0.65ex,scale=0.5]
\filldraw[black] (0,0) circle (4pt);
    \draw[->] (-0.6,-0.6) -- (0.6,0.6);
    \draw[<-] (-0.6,0.6) -- (0.6,-0.6);
    \node at (-1.3,1.3) (a1) {$\pm\frac12$};
       \node at (1.3,-1.3) (a2) {$\mp\frac12$};
           \node at (1.3,1.3) (a3) {$\mp\frac12$};
    \node at (-1.3,-1.3) (a4) {$\pm\frac12$};
\end{tikzpicture}=f(\theta)x^{\pm1}(q-q^{-1})\ .
\label{wea2}
}

The braiding relation \eqref{bru} is automatically satisfied because the Hecke algebra relation \eqref{eq:Hecke} implies 
\EQ{
\check R(x)\check R(x^{-1})=(xq-x^{-1}q^{-1})(x^{-1}q-xq^{-1})\ ,
}
as long as the scalar factor obeys
\EQ{
f(\theta)f(-\theta)=\frac1{(xq-x^{-1}q^{-1})(x^{-1}q-xq^{-1})}\ .
\label{re2}
}

Unitarity then follows if the S-matrix is Hermitian analytic
\EQ{
S_I(\theta)=S_I(-\theta^*)^*\ ,\qquad S_T(\theta)=S_T(-\theta^*)^*\ ,\qquad S_R^\pm(\theta)=S_R^\pm(-\theta^*)^*\ ,
}
providing the scalar factor satisfies 
\EQ{
f(\theta^*)^*=-f(-\theta)\ .
\label{ha1}
}

Crossing symmetry requires that either
\EQ{
x=q^{-\theta/(i\pi)}\qquad\text{or}\qquad x=(-q)^{-\theta/(i\pi)}\ .
\label{y77}
}
It turns out that the resulting S-matrices are physically equivalent and so we choose the former. However, with this choice some extra factors of $-1$ appear in the crossing symmetry relation and charge conjugation operator; however, these are unobservable.\footnote{See the discussion in Appendix C of \cite{Bernard:1990ys} for details.} Crossing symmetry implies
\EQ{
S_I(\theta)=S_T(i\pi-\theta)\ ,\qquad S_R^+(\theta)=q^{-1}S_R^-(i\pi-\theta)\ ,
\label{ha2}
}
where the charge conjugation operator acts as 
\EQ{
{\cal C}\ket{\pm\tfrac12,\theta}=\pm i q^{\mp 1/2}\ket{\mp\tfrac12,\theta}
\label{phicross}
}
and there is a further constraint on the scalar factor:
\EQ{
f(\theta)=f(i\pi-\theta)\ .
\label{re1}
}

In addition, if the theory is parity symmetric then one has an additional constraint on the reflection amplitudes\footnote{Since parity flips the spatial coordinate, the ordering of particles is interchanged. Parity also flips momenta $p_i = m \sinh \theta_i$  and so sends $\theta_i \rightarrow - \theta_i$.  However the rapidity $\theta$ in the S-matrix is the rapidity difference of particles and so remains unchanged under the combined action of flipping the order and momenta of individual particles.}
\EQ{
S_R^\pm(\theta)=S_R^\mp(\theta)\ .
}

For the sine-Gordon theory, the S-matrix was originally constructed in the seminal work of Zamolodchikov and Zamolodchikov \cite{Zamolodchikov:1978xm}. In this case, with the former choice in \eqref{y77}, we have
\EQ{
q=\exp[-i\pi/\gamma']\ ,\qquad x(\theta)=\exp[\theta/\gamma']\ .
}
However, there is a problem: the S-matrix as written is not Hermitian analytic:
the reflection amplitudes are non-compliant because they satisfy
\EQ{
S_R^\pm(\theta^*)^*=S_R^\mp(-\theta)\ ,
\label{vha}
}
clearly violating \eqref{ha2}.

Hermitian analyticity can, however, be restored by a simple rapidity-dependent transformation on the states of the form \cite{LeClair:1989wy,Bernard:1990cw,Bernard:1990ys}
\EQ{\label{eq:gradationshift}
\ket{\pm\tfrac12,\theta}\longrightarrow x(\theta)^{\pm1/2}\ket{\pm\tfrac12,\theta}\ .
}
This transformation removes the factors of $x^{\pm1}$ from the reflection amplitudes and restores Hermitian analyticity.\footnote{To ensure crossing symmetry charge conjugation needs to be modified so that ${\cal C} \ket{\pm \frac12} =\pm \ket{\mp \frac12}$, in agreement with the original construction of \cite{Zamolodchikov:1978xm} but with the additional factors of $\pm$ needed for the choice made for $q$ explained in \cite{Bernard:1990ys}.}
It has an algebraic interpretation of moving from the homogeneous to the principal gradation of the affine algebra $\widehat{\msu(2)}$.\footnote{In more detail the change of grade can be achieved by a re-scaling and  conjugation on the loop algebra $a(z)\to Ua(z^2)U^{-1}$ where $U=z^{i \sqrt2T^3}$. Then with the identification $z^2=x$, the conjugation is precisely the transformation \eqref{eq:gradationshift} on states.} The transmission and identical amplitudes are insensitive to this change whereas in the principal gradation the reflection amplitudes become 
\be
S_R^+(\theta) = S_R^-(\theta) = f(\theta) (q - q^{-1}) \ , 
\ee
such that  the resulting S-matrix now also describes a parity symmetric theory.  Due to the change in gradation, the S-matrix is now invariant under the affine quantum group in principal gradation $\mathscr U_q(\widehat{\msu(2)}_p)$.

The result S-matrix is precisely the S-matrix of the solitons of the sine-Gordon theory once we specify the scalar function $f(\theta)$. This is not determined uniquely by the conditions \eqref{re2}, \eqref{ha1} and \eqref{re1}. However, we can invoke the concept of {\it minimality\/} meaning that the solution has the minimal number of poles on the {\it physical strip:} $0<\IM\theta<\pi$. The significance of this is that poles on the physical strip along the imaginary axis are usually interpreted in terms of bound states propagating in either the direct or crossed channels.\footnote{There are also double poles which are  explicable as anomalous thresholds.} The minimal expression can be written in various ways, for example as
\EQ{
f(\theta,\gamma')=\frac1{2\pi i}\prod_{n=1}^\infty\frac{\Gamma(\frac{2n}{\gamma'}+\frac{i\theta}{\pi\gamma'})\Gamma(1+\frac{2n-2}{\gamma'}+\frac{i\theta}{\pi\gamma'})\Gamma(\frac{2n-1}{\gamma'}-\frac{i\theta}{\pi\gamma'})\Gamma(1+\frac{2n-3}{\gamma'}-\frac{i\theta}{\pi\gamma'})} {\Gamma(\frac{2n+1}{\gamma'}+\frac{i\theta}{\pi\gamma'})\Gamma(1+\frac{2n-1}{\gamma'}+\frac{i\theta}{\pi\gamma'})\Gamma(\frac{2n}{\gamma'}-\frac{i\theta}{\pi\gamma'})\Gamma(1+\frac{2n-2}{\gamma'}-\frac{i\theta}{\pi\gamma'})}\ .
\label{sgs}
}
It is simple to show that this solves the conditions by computing its divisor. The other important condition that this expression satisfies is the Hermitian analyticity condition \eqref{ha1}. Another way to write the results that will be useful later is as the integral expression (valid for $\gamma'>2$)
\EQ{
f(\theta,\gamma')=\frac1{q-q^{-1}}\exp\Big\{2\int_0^\infty\frac {dw}w\,\frac{\cosh[\pi w(\gamma'-2)/2]\sin[w(i\pi-\theta)/2]\sin[w\theta/2]}{\cosh[\pi w/2]\sinh[\pi w\gamma'/2]}\Big\}\ .
\label{rr2}
}

The sine-Gordon S-matrix also has an RSOS cousin, the restricted sine-Gordon S-matrix \cite{LeClair:1989wy,Bernard:1990ys,Bernard:1990cw} which is associated to case when $q$ is a root-of-unity. Details of this will emerge in section \ref{s5.3}. 

\subsection{Quantum group S-matrix: $\boldsymbol q$ real}\label{s5.2}

Given that the YB and XXZ in the regime $\xi>\lambda$ have a quantum group parameter $q=\exp[-\pi/\sigma]$ that is real, implies that we also need an S-matrix that will be a close cousin of the sine-Gordon S-matrix but with this real value of $q$. On top of this, since the resulting theories have a cyclic RG behaviour, heuristic arguments suggest that the S-matrix should have a periodicity in real rapidity \cite{Leclair:2003xj}:\footnote{We follow the convention of \cite{Leclair:2003xj} and allow the S-matrix to actually change up to some minus signs over a period. Note that S-matrices with a real periodicity in rapidity cannot have bound states but can have an infinite set of resonance poles \cite{Leclair:2003xj,Mussardo:1999ee}.}
\EQ{
S(\theta+\pi\sigma)=S(\theta)\ .
}
This periodicity requires more than a simple analytic continuation of couplings.  Such an S-matrix was constructed in \cite{Leclair:2003xj} built on the same quantum group $R$-matrix as the sine-Gordon S-matrix but now with real $q$. Crossing symmetry now requires that
\EQ{
x=\exp[-i\theta/\sigma]\ .
}

The $R$-matrix now has a periodicity under shifts $\theta\to\theta+\pi\sigma$ (more precisely up to some minus signs). This periodicity can be inherited by the S-matrix if the scalar factor is such that
\EQ{
f(\theta+\pi\sigma;\sigma)=f(\theta;\sigma)\ .
\label{rp1}
}
The situation with Hermitian analyticity is different from the real $\gamma'$ regime: both the S-matrix in the principal {\it and\/} homogeneous gradations are Hermitian analytic as long as the scalar factor satisfies \eqref{ha1}. In principle grade, the two refection amplitudes $S_R^\pm$ are equal:
\EQ{
S_R^\pm(\theta)\equiv S_R(\theta)=f(\theta;\sigma)(q-q^{-1})\ .
}
While in homogeneous grade, the two refection amplitudes differ:
\EQ{
S_R^\pm(\theta)=f(\theta;\sigma)x^{\pm1}(q-q^{-1})\ .
}

To complete the construction we must specify the scalar factor.
Note that simply taking the analytic continuation of the sine-Gordon scalar factor \eqref{sgs} from $\gamma'\to i\sigma$ would not have the requisite periodicity \eqref{rp1} or satisfy the Hermitian analyticity constraint \eqref{ha1}. On the contrary, the minimal solution to the constraints can be written as the convergent product \cite{Leclair:2003xj}
\EQ{
f(\theta;\sigma)=q\prod_{n=1}^\infty\frac{(1-q^{4n}x^{-2})(1-q^{4n+2}x^2)}{(1-q^{4n}x^2)(1-q^{4n-2}x^{-2})}\ .
}
Note that this immediately satisfies \eqref{ha1} and is manifestly periodic under $\theta\to\theta+\pi\sigma$.

So there are two consistent S-matrices $S_h(\theta;\sigma)$ and $S_p(\theta;\sigma)$, associated to the homogeneous and principal gradations, respectively. It is important that the S-matrix that uses the homogeneous gradation of the affine quantum group, breaks parity $S_R^+(\theta)\neq S_R^-(\theta)$, whereas the principal gradation case preserves parity.

The other important point to emphasize here is that when $q$ is real, the S-matrix associated to the affine quantum group $\mathscr U_q(\widehat{\msu(2)})$ automatically has the periodicity in real rapidity that  matches the heuristic proposal of \cite{Leclair:2003xj} that theories with cyclic RG behaviour should have just such a periodicity at high centre-of-mass energy. But note that the S-matrix goes beyond this because it has the periodicity for {\it any\/} centre-of-mass energy.

\subsection{The RSOS S-matrix}\label{s5.3}

In order construct our S-matrices we will also need a piece to handle the kink quantum numbers of the states. 
This is precisely the RSOS kink S-matrix of the restricted sine-Gordon theory \cite{LeClair:1989wy,Bernard:1990ys,Bernard:1990cw}. It is built out of a solution of the Yang-Baxter Equation, or more precisely the {\it star-triangle relation\/}, that plays the role of  Boltzmann weights in an Interaction Round a Face (IFR) statistical model, e.g.~see \cite{Jimbo:1987ra}.

In the IRF S-matrix, the states are kinks $K_{ab}(\theta)$ and states are labelled by the vacua $a,b$ on either side.  The vacua (the local heights of the statistical model) are associated to representations of $\mathscr U_q(\msu(2))$ so to spins $a,b,\ldots\in\{0,\frac12,1,\frac32,\ldots\}$. When $q$, the quantum group a parameter is a root of unit, 
\EQ{
q=\exp\big[-i\pi/(k+2)\big]\ ,
}
there is a restricted model, where the spins are restricted to lie in the set of integrable representations of level $\leq k$, so $a,b,\ldots\in\{0,\frac12,1,\dots,\frac k2\}$. A basis of states in the Hilbert space with $N$ kinks is labelled by a sequence $\{a_{N+1},a_N,\ldots,a_1\}$, which has the interpretation of a {\it fusion path\/}, so the spin $a_{j+1}$ representation must appear in the tensor product of the $a_j$ representation with the spin $\frac12$ representation (truncated by the level restriction). This means that there is an adajency condition $a_{j+1}=a_j\pm\frac12$. 

The analogue of the $R$-matrix, is an intertwiner $W$ between  2-kink states \cite{Jimbo:1987ra}:
\EQ{
\ket{K_{ab}(\theta_1)K_{bc}(\theta_2)}\longrightarrow \sum_dW\left.\left.\left(\hspace{-0.2cm}\raisebox{-14.5pt}{\begin{tikzpicture}[scale=0.32]
\node at (0,1.1) {$d$};
\node at (0,-1.1) {$b$};
\node at (1.2,0) {$c$};
\node at (-1.2,0) {$a$};
\end{tikzpicture}}\hspace{-0.15cm}\right.\right| u\right)\ket{K_{ad}(\theta_2)K_{dc}(\theta_1)}\ ,
}
where $u=\theta/(i\pi)$ and $\theta=\theta_1-\theta_2$. These intertwiners satisfy the star triangle relation \cite{Jimbo:1987ra}.

The solution of the star triangle relation $W(u)$ is the raw fodder from which one  fashions the RSOS S-matrix for kinks states. There are 3 basic types of non-vanishing elements that take the form
\EQ{
W\left.\left.\left(\hspace{-0.2cm}\raisebox{-21.5pt}{\begin{tikzpicture}[scale=0.5]
\node at (0,1) {\footnotesize $a\pm\frac12$};
\node at (0,-1) {\footnotesize $a\pm\frac12$};
\node at (1.3,0) {\footnotesize $a$};
\node at (-1.1,0) {\footnotesize $a\pm1$};
\end{tikzpicture}}\hspace{-0.15cm}\right.\right|u\right)&=\frac{[1-u]}{[1]}\ ,\\[5pt]
W\left.\left.\left(\hspace{-0.2cm}\raisebox{-21.5pt}{\begin{tikzpicture}[scale=0.5]
\node at (0,1) {\footnotesize $a\pm\frac12$};
\node at (0,-1) {\footnotesize $a\pm\frac12$};
\node at (1.5,0) {\footnotesize $a$};
\node at (-1.5,0) {\footnotesize $a$};
\end{tikzpicture}}\hspace{-0.15cm}\right.\right|u\right)&=\frac{[\pm(2a+1)+u]}{[\pm(2a+1)]}\ ,\\[5pt]
W\left.\left.\left(\hspace{-0.2cm}\raisebox{-21.5pt}{\begin{tikzpicture}[scale=0.5]
\node at (0,1) {\footnotesize $a\mp\frac12$};
\node at (0,-1) {\footnotesize $a\pm\frac12$};
\node at (1.5,0) {\footnotesize $a$};
\node at (-1.5,0) {\footnotesize $a$};
\end{tikzpicture}}\hspace{-0.15cm}\right.\right|u\right)&=\frac{[u]}{[1]}\cdot\frac{\sqrt{[2a+2][2a]}}{[2a+1]}\ ,
\label{wwd}
}
where we have defined
\EQ{
[u]=\sin\big(\pi u/(k+2)\big)\ .
}

The $W$ intertwiner satisfies some identities that are important for the S-matrix that we going to build \cite{Jimbo:1987ra}:
(i) the {\it initial condition\/}
\EQ{
W\left.\left.\left(\hspace{-0.2cm}\raisebox{-14.5pt}{\begin{tikzpicture}[scale=0.32]
\node at (0,1) {$d$};
\node at (0,-1) {$b$};
\node at (1.2,0) {$c$};
\node at (-1.2,0) {$a$};
\end{tikzpicture}}\hspace{-0.15cm}\right.\right|0\right)=\delta_{bd}\ ;
}
(ii) {\it rotational symmetry\/}
\EQ{
W\left.\left.\left(\hspace{-0.2cm}\raisebox{-14.5pt}{\begin{tikzpicture}[scale=0.32]
\node at (0,1) {$d$};
\node at (0,-1) {$b$};
\node at (1.2,0) {$c$};
\node at (-1.2,0) {$a$};
\end{tikzpicture}}\hspace{-0.15cm}\right.\right|1-u\right)=\sqrt{\frac{[2b+1][2d+1]}{[2a+1][2c+1]}}W\left.\left.\left(\hspace{-0.2cm}\raisebox{-13.5pt}{\begin{tikzpicture}[scale=0.32]
\node at (0,1) {$c$};
\node at (0,-1.2) {$a$};
\node at (1.2,0) {$b$};
\node at (-1.2,0) {$d$};
\end{tikzpicture}}\hspace{-0.15cm}\right.\right|u\right)\ ;
}
and (iii) {\it inversion relation\/}
\EQ{
\sum_d W\left.\left.\left(\hspace{-0.2cm}\raisebox{-14.5pt}{\begin{tikzpicture}[scale=0.32]
\node at (0,1) {$d$};
\node at (0,-1) {$b$};
\node at (1.2,0) {$c$};
\node at (-1.2,0) {$a$};
\end{tikzpicture}}\hspace{-0.15cm}\right.\right|u\right)
W\left.\left.\left(\hspace{-0.2cm}\raisebox{-14.5pt}{\begin{tikzpicture}[scale=0.32]
\node at (0,1) {$e$};
\node at (0,-1) {$d$};
\node at (1.2,0) {$c$};
\node at (-1.2,0) {$a$};
\end{tikzpicture}}\hspace{-0.15cm}\right.\right|-u\right)=\frac{[1-u][1+u]}{[1]^2}\delta_{be}\ .
}
The alert reader will recognize that the rotational symmetry and inversion relation as proto-identities for crossing symmetry and braiding unitarity, respectively.

When $k$ is generic (i.e. not an integer), the local heights $a,b,\ldots$ are valued in $\frac12\mathbb Z$ and the Boltzmann weights $W(u)$ define the SOS statistical model. However, when $k$ is an integer there is consistent restriction of the local heights to the finite set $\{0,\frac12,1,\ldots,\frac k2\}$. The restriction is consistent because
$[0]=[k+2]=0$ so consequently $W(u)$ cannot propagate a kink state with admissible local heights $\ket{K_{ab}(\theta_1)K_{bc}(\theta_2)}$ with $a,b,c\in\{0,\frac12,1,\ldots,\frac k2\}$ into one with an inadmissible local height $\ket{K_{ad}(\theta_2)K_{dc}(\theta_1)}$ with $d\not\in\{0,\frac12,1,\ldots,\frac k2\}$, in practice $d=0$ or $\frac k2+1$, due to the adjacency condition. This is guaranteed if $[0]=[k+2]=0$.  

In order to make a consistent S-matrix,
\EQ{
S_\text{RSOS}(\theta;k)=v(\theta)W\left.\left.\left(\hspace{-0.2cm}\raisebox{-14.5pt}{\begin{tikzpicture}[scale=0.32]
\node at (0,1) {$d$};
\node at (0,-1) {$b$};
\node at (1.2,0) {$c$};
\node at (-1.2,0) {$a$};
\end{tikzpicture}}\hspace{-0.15cm}\right.\right|u(\theta)\right)\ ,
}
one has to construct a suitable scalar factor $v(\theta)$ in order that the S-matrix is unitary and crossing symmetric. The scalar factor must satisfy
\EQ{
v(\theta)=v(i\pi-\theta)\ ,\qquad v(\theta)v(-\theta)=\frac{\sin^2(\pi/(k+2))}{\sin((\pi+i\theta)/(k+2))\sin((\pi-i\theta)/(k+2))}\ .
}

One can readily verify that the solution to these conditions can be expressed in terms of the usual sine-Gordon scalar factor in \eqref{rr2} with $\gamma'=k+2$, up to a constant factor:
\EQ{
v(\theta;k)=(q-q^{-1})f(\theta;k+2)\ .
}
where $f(\theta;\gamma')$ is defined in \eqref{sgs}.

The RSOS kink S-matrix has a good limit $k\to\infty$, the SOS limit,  as long as the local heights are suitably shifted, $a\to \frac k4+a$, etc, before the limit is taken. So the idea is that one takes the local heights well away from the end points $a=0$ and $a=\frac k2$ as $k\to\infty$. In that limit, one can easily verify that the S-matrix becomes identical to the rational $\SU(2)$ S-matrix with a simple mapping between the kinks of the SOS picture and states of the spin $\frac12$ representation:
\EQ{
K_{a+\frac12.a}(\theta) \longleftrightarrow \ket{\uparrow;\theta}\ ,\qquad K_{a-\frac12,a}(\theta) \longleftrightarrow \ket{\downarrow;\theta}\ .
}
This is an IRF-to-vertex transformation which relies on the fact that 
the $N$-kink Hilbert space of unrestricted paths of length $N$ $\{a_{N+1},a_N,\ldots,a_1\}$, is isomorphic to the $N$ spin $\frac12$ particle Hilbert space for a fixed $a_1$; e.g.
\EQ{
 \{a+1,a+\tfrac32,a+1,a+\tfrac32,a+1,a+\tfrac12,a\}\longleftrightarrow \ket{\downarrow\uparrow\downarrow\uparrow\uparrow\uparrow}\ ,
 }
 etc.

Finally, we can compare our S-matrices by writing down an integral representations of the identical particle amplitude, which for the RSOS case means 
\EQ{
\ket{K_{a\pm1,a\pm\frac12}(\theta_1)K_{a\pm\frac12,a}(\theta_2)}\to \ket{K_{a\pm1,a\pm\frac12}(\theta_2)K_{a\pm\frac12,a}(\theta_1)}\ .
} 
Note that this particular amplitude does not depend on the right vacuum $a$.

For the $q$ a complex phase---the sine-Gordon case---we have
\EQ{
S_I(\theta;\gamma')=\exp\Big\{i\int_0^\infty\frac{dw}w\,\frac{\sin[w\theta]\sinh[\pi w(\gamma'-1)/2]}{\cosh[\pi w/2]\sinh[\pi w\gamma'/2]}\Big\}\ .
\label{g4r}
}
For the case $q$ real, the S-matrix of \cite{Leclair:2003xj}, we have
\EQ{
S_I(\theta;\sigma)=\exp\Big\{i\theta/\sigma+i\sum_{n=1}^\infty\frac2n\cdot\frac{\sin[2n\theta/\sigma]}{1+\exp[2\pi n/\sigma]}\Big\}\ .
}
Finally for the RSOS case just constructed
\EQ{
S_{I,\text{RSOS}}(\theta;\sigma,k)=\exp\Big\{i\int_0^\infty\frac{dw}w\,\frac{\sin[w\theta]\sinh[\pi w(k+1)/2]}{\cosh[\pi w/2]\sinh[\pi w(k+2)/2]}\Big\}\ ,
}
which is simply \eqref{g4r} with $\gamma'\to k+2$.

\pgfdeclareimage[interpolate=true,width=8cm]{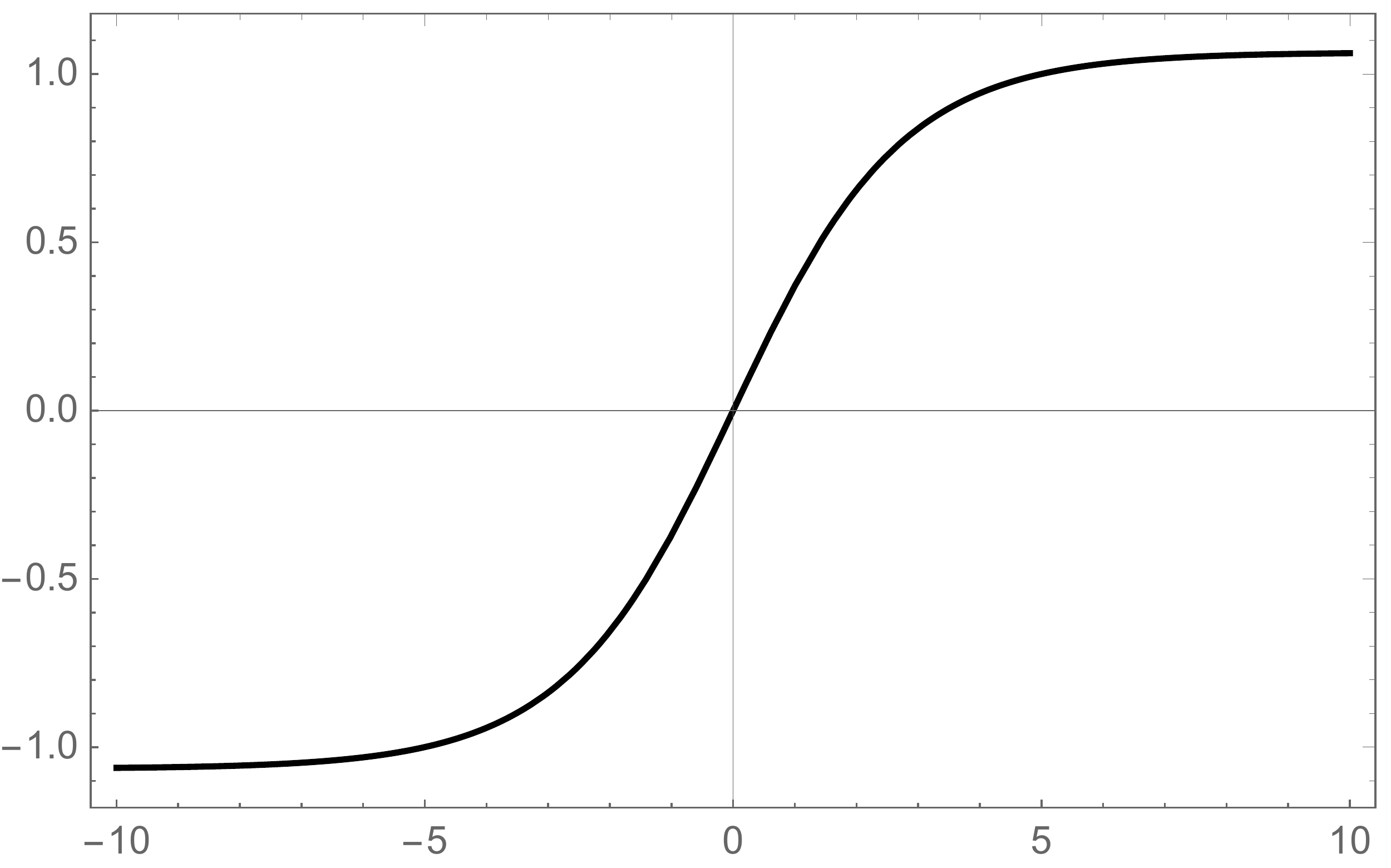}{fs3}
\begin{figure}
\begin{center}
\begin{tikzpicture}[scale=1]
\pgftext[at=\pgfpoint{0cm}{0cm},left,base]{\pgfuseimage{fs3}} 
\node[rotate=90] at (-0.6,2.5) {\footnotesize $-i\log S_I(\theta;\gamma')$};
\node at (4,-0.4) {$\theta$};
\end{tikzpicture}
\caption{\footnotesize The identical particle scattering phase as a function of the rapidity for some indicative value of $\gamma'$. The key feature is that for large $\theta$ the amplitude saturates.}
\label{f11}
\end{center}
\end{figure}
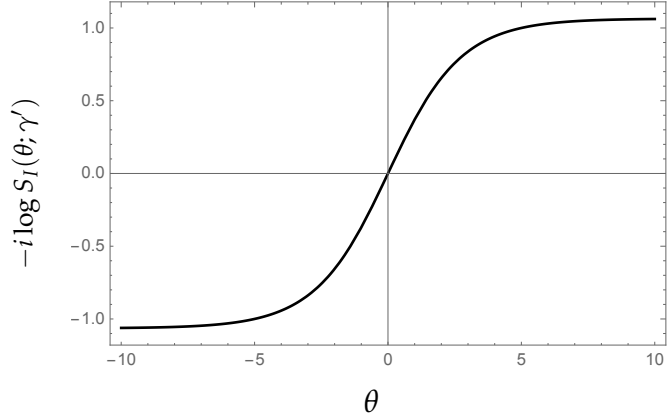

\subsection{High energy limit}\label{s5.4}

The final information we will need when we establish our S-matrix conjectures is the high centre-of-mass energy limit of the trigonometric S-matrices. This is just the large rapidity limit, i.e.~the limit of large $x$ defined in \eqref{y77} . In order to take the limit, we focus on the identical particle amplitude $S_I$ which can be written (by rearranging the arguments of the gamma functions in \eqref{sgs}) as
\EQ{
S_I(\theta;\gamma')&=\prod_{j=1}^\infty \frac{\Gamma(\tfrac{i\theta}{2\pi}+\tfrac j2\gamma')\Gamma(\tfrac{i\theta}{2\pi}+1+\tfrac{j-1}2\gamma')}{\Gamma(-\tfrac{i\theta}{2\pi}+\tfrac j2\gamma')\Gamma(-\tfrac{i\theta}{2\pi}+1+\tfrac{j-1}2\gamma')}\\
&\qquad\qquad \times \frac{\Gamma(-\tfrac{i\theta}{2\pi}+\tfrac12+\tfrac j2\gamma')\Gamma(-\tfrac{i\theta}{2\pi}+\tfrac12+\tfrac{j-1}2\gamma')}{\Gamma(\tfrac{i\theta}{2\pi}+\tfrac12+\tfrac j2\gamma')\Gamma(\tfrac{i\theta}{2\pi}+\tfrac12+\tfrac{j-1}2\gamma')}\ .
\label{poo}
}
Note that this amplitude is also valid in the RSOS version of the S-matrix with $\gamma'\to k+2$.

The amplitude is a phase which we plot in Fig.~\ref{f11}. The important point is that for large enough $\theta$ the amplitude saturates. In order calculate the asymptotic value we simply apply Stirling's formula to the expression above:
\EQ{
-i\log S_I(\theta;\gamma') & \longrightarrow \frac{\theta}{4\pi}(\gamma'-1)\sum_{j=1}^\infty\frac1{(2\gamma' j)^2+(\theta/(2\pi))^2}+\cdots\\
&=\frac{\pi}2(1-\gamma')+\cdots\ .
}
This means that the while S-matrix has a very simple limit proportional to the Hecke algebra generator
\EQ{
S(\theta;\gamma')\overset{\theta\gg1}{\xrightarrow{\hspace*{1.3cm}}} e^{i\pi(1+\gamma')/2}T^{-1}\ .
}
Note that the RSOS kink S-matrix also has such a universal high energy limit, where now $T^{-1}$ is realized in the kink Hilbert space.

\subsection{The S-matrix proposals}\label{s5.4}

In this section, based on all the information and constraints, we make our proposals for the S-matrices of the lambda and sigma models.

We begin with the XXZ lambda model in the regime with $\gamma'\in\mathbb R$, i.e.~the quantum group parameter a complex phase. Our proposal is that the S-matrix in this regime, is precisely the fractional sine-Gordon S-matrix \eqref{h90} proposed by Bernard and LeClair \cite{Bernard:1990ti}. The theory in this regime has a pair of affine quantum group symmetries with $q=\exp[-i\pi/\gamma']$, for the sine-Gordon factor, and $q=\exp[-i\pi/(k+2)]$ for the RSOS factor.

The S-matrices for the sigma model follows in the limit $k\to\infty$ and a non-abelian T-duality which has the effect of replacing the RSOS S-matrix piece  with the rational $\SU(2)$ S-matrix as shown in \eqref{x99} and one recovered the S-matrix of the anisotropic XXZ sigma model in \eqref{w8e}.

Now we turn to the YB lambda model and the XXZ model in the regime $\gamma'=i\sigma$, $\sigma\in\mathbb R$, i.e.~where the quantum group parameter $q=\exp[-\pi/\sigma]$ is real. In these case the RG flows are cyclic. This suggest that the S-matrices are based on the pieces
$S_h(\theta;\sigma)$  and $S_p(\theta;\sigma)$ constructed in section \ref{s5.2}. There is also a natural explanation for the existence of the two distinct S-matrices based on the gradation because the YB lambda model is not  parity symmetric and this matches the S-matrix for the homogeneous gradation. Correspondingly the principal gradation S-matrix is parity  preserving as is the XXZ model. 

To make a complete S-matrix we need to consider an appropriate RSOS kink S-matrix factor. The the only choice consistent with the sigma model and the classical symmetries is the RSOS S-matrix piece $S_\text{RSOS}(\theta;k)$. However, this S-matrix does not have the periodicity $\theta\to\theta+\pi\sigma$.
The resolution is here is that the periodicity is only expected to appear in the limit of large centre-of-mass energy and we have shown in section \ref{s5.4} that the trigonometric S-matrix become constant at high energy. So the heuristic requirement that the S-matrices of theories with cyclic RG behaviour should have a periodicity in rapidity at high energy is actually satisfied.

Hence, we make our conjectures; for the YB lambda model
\EQ{
S_{\lambda\text{-YB}}(\theta)=S_\text{RSOS}(\theta;k)\otimes S_h(\theta;\sigma)\ ,
}
while for the XXZ lambda model
\EQ{
S_{\lambda\text{-XXZ}}(\theta)=S_\text{RSOS}(\theta;k)\otimes S_p(\theta;\sigma)\ .
\label{x56}
}

The sigma model limit, involves taking $k\to\infty$ along with an IRF-to-vertex transformation, 
\EQ{
S_{\sigma\text{-YB}}(\theta)&=S_{\SU(2)_L}(\theta)\otimes S_h(\theta;\sigma)\ ,\\ 
S_{\sigma\text{-XXZ}}(\theta)&=S_{\SU(2)_L}(\theta;\sigma)\otimes S_p(\theta;\sigma)\ ,
}
respectively. These S-matrices exhibit the Yangian $\mathscr Y(\msu(2))$ symmetry and also have the periodicity in rapidity $\theta\to\theta+\pi\sigma$ at high energy.

\section{Discussion}

In this work we have considered the deformations of the $\SU(2)$ PCM that preserve integrability. The class of deformations focused on, preserved an $\SU(2)$ symmetry and so there are associated lambda models. We showed that the lambda models also have affine quantum group symmetries realized at the classical Poison bracket level. The are many questions remaining. In particular, for the YB deformations and anisotropic ones with $\beta>\alpha$ (or $\xi>\lambda$ for the associated lambda model), the RG flow follows a cycle in coupling constant space. So these theories have a mass gap but no fixed point in the UV to define a continuum limit. So the main question is: is the UV of these theories well defined? There are two pieces of evidence to suggest that these theories actually are only defined with an explicit UV cut off of the order of the mass scale of the particle states. 

The first, described in \cite{Leclair:2003xj} for the case $k=1$, comes from defining the QFT as the continuum limit of a spin chain. The anisotropic XXZlambda models with $\xi<\lambda$, so with $q$ in \eqref{u7a} a complex phase, can be regularized by the XXZ Heisenberg spin chain \cite{forthcoming} with spins of angular momentum $j=\frac k2$ and with spin chain anisotropy 
\EQ{
\Delta=\cos\frac\pi{k+\gamma'}\ ,
}
where $\gamma'$ is the RG invariant \eqref{rgi}. The spin chain in this regime is critical and consequently it is possible to take a continuum limit. The physical excitations and their S-matrix agree precisely with our conjectured S-matrix, the fractional sine-Gordon S-matrix in \eqref{h90}. Now if we try a similar spin chain description of the $\xi>\lambda$ case, then the XXZ spin chain now lies in the $\Delta<-1$ regime. In this regime the spin chain has a mass gap and so there is no way to take a continuum limit. Even so, we shall show in  \cite{forthcoming}, that the excitations have an S-matrix that is a close relative of the S-matrix \eqref{x56}  It is possible to create a hierarchy between the inverse  lattice spacing and excitation mass only in the limit of large $\sigma$. So this suggests that the RG cycle is never actually traversed in the UV before the UV cut off is reached. 

The second piece of evidence, again for the case $k=1$ for the anisotropic XXZ lambda model in the cyclic RG regime, is presented in \cite{LeClair:2003hj}. The idea is to use finite size effects to compute the effective central charge. It is shown that for the case when the theory has a mass gap, the relevant case here, the finite-size effects do indeed have a periodic behaviour consistent with the beta function analysis but in the deep UV the finite-size central charge either has a singularity or is ill defined in the very deep UV. Again this suggests that in the cyclic RG regime, the theories only make sense with an explicit UV cut off.

The other issue which is interesting to consider is how these issues play out in larger groups. We have already pointed out that the anisotropic models are special to $\SU(2)$ and they do not appear to admit generalizations to an arbitrary Lie group. However, the Yang-Baxter deformation do lift to an arbitrary group and one can speculate that the sigma and lambda models once again have a cyclic RG behaviour. We show this is the case in  \cite{forthcoming}. We go on to show that there is a natural conjecture for the S-matrix which is rather novel. For the case $\SU(N)$, it is related to the S-matrix constructed in \cite{Hollowood:1992sy} but like the S-matrix $S(\theta;\sigma)$ considered here is periodic in rapidity. What is novel about the resulting S-matrix is that it exhibits an infinite set of unstable resonance poles thus providing an example of the ``Russian Doll" phenomena described in \cite{LeClair:2003hj}. Unlike the $\SU(2)$ example described there, the S-matrix we construct satisfies all the S-matrix axioms including hermitian analyticity.

Finally there is a generalization of the anisotropic models that we have mentioned in the introduction, namely the XYZ model. The lambda model of this should have an S-matrix that is related to the elliptic S-matrix of Zamolodchikov \cite{Zamolodchikov:1979ba}.

\section*{Acknowledgements}

\noindent
CA and DP are supported by STFC studentships.  TJH is supported in part by the STFC grant ST/L000369/1. 
 DCT is supported by a Royal Society University Research Fellowship {\em Generalised Dualities in String Theory and Holography} UF 150185. 
We would like to thank  Saskia Demulder, Kostas Sfetsos, Graham Shore, Kostas Siampos and Benoit Vicedo for useful discussions and Arkady Tseytlin for interesting correspondence.

\appendix
\appendixpage

\section{Lambda Spacetimes}\label{a1}
 
The lambda theories can viewed as sigma models with target spaces of the following form
 \be
\begin{aligned}
ds^2 &= \frac{k}{A_0} \left( A_1 \,d\phi^2 + A_2 \,d\psi^2 + A_3 d\theta^2 + A_4 \,d\phi \,d\psi \right) \ ,   \\
H_3 &= k \frac{A_5 }{A_0^2}  \,d\phi\wedge d\psi \wedge d\theta \ , \\
\Phi &= -\frac{1}{2} \log A_0 = - \frac{1}{2} \log \det   (\BOmega  - {\textrm{Ad}}_{\cal F} )  \  , \quad A_i = A_i(\phi, \psi) \ . 
\end{aligned}
\ee
The non-trivial dilaton is produced as a result of the determinant in the path integral arising from performing the Gaussian integration on the non-propagating gauge fields.
Explicitly one has for the YB lambda model (with $c^{-1} = \lambda^3(1+\eta^2)$) \cite{Sfetsos:2015nya}:
\be
\begin{aligned}
A_0 &=c (\lambda -1) \left(4 \lambda  \left(C_{\phi }-\eta  C_{\psi } S_{\phi }\right)
   \left(\eta  \lambda  C_{\psi } S_{\phi }+C_{\phi }\right)-(\lambda +1)^2\right) \\
 A_1 &=   c \left((\lambda +1) \left(-2 \lambda  C_{2 \phi } \left(\eta ^2 \lambda
   +1\right)+\left(2 \eta ^2+1\right) \lambda ^2+1\right)-2 \eta  \lambda  \left(\lambda
   ^2-1\right) C_{\psi } S_{2 \phi }\right)\\
   A_2 &= c (\lambda -1)^2 (\lambda +1) S_{\phi }^2\\
   A_3 &= c (\lambda -1)^2 (\lambda +1) S_{\psi }^2 S_{\phi }^2\\
   A_4 & =\frac{4 \eta  \left(\lambda ^2-1\right) S_{\psi } S_{\phi }^2}{\left(\eta ^2+1\right)
   \lambda ^2} \\
   A_5 &=2 c^2 (\lambda -1)^2 S_{\phi }^2 \left[S_{\psi } \left\{\lambda ^2 C_{2 \phi } \left(2
   \eta ^4 \lambda ^2-3 \eta ^2 \left(\lambda ^2+1\right)-8\right)   \right. \right. \\ &\qquad \left.\left.   + \lambda 
   \left(\lambda  \left(-2 \eta ^4 \lambda ^2-\eta ^2 ((\lambda -8) \lambda +1)+2
   (\lambda  (\lambda +2)-2)\right)+4\right)+2\right\} \right. \\& \qquad \left. +2 \eta  \lambda ^2 \left(2
   (\lambda -1) \left(\eta ^2 \lambda -1\right) S_{2 \psi } S_{2 \phi }-\eta  \left(2
   \eta ^2 \lambda ^2+\lambda ^2+1\right) S_{3 \psi } S_{\phi }^2\right)\right]
   \ . 
\end{aligned}   
\ee 
For the XXZ lambda model we find (with $c= \xi^{-2} \lambda^{-1}$ ):
\be
\begin{aligned}
A_0 &=  -c \left((\xi +1) \left(C_{2 \psi } (\lambda -\xi )+\lambda  \xi -1\right)+2 C_{\phi }^2
   \left((\xi +1) C_{\psi }^2 (\xi -\lambda )-(\xi -1) (\lambda +\xi )\right)\right)\\
   A_1&=     c \left(2 C_{\phi }^2 \left((\xi -1) C_{\psi }^2 (\xi -\lambda )-(\xi +1) (\lambda +\xi
   )\right)+(\lambda +1) (\xi +1)^2\right) \\
A_2&=c (\xi -1) S_{\phi }^2 \left(C_{2 \psi } (\lambda -\xi )+\lambda  \xi -1\right)\\ 
A_3 &=c (\lambda -1) \left(\xi ^2-1\right) S_{\psi }^2 S_{\phi }^2\\
A_4&= c (\xi -1) (\lambda -\xi ) S_{2 \psi } S_{2 \phi }\\ 
A_5&= -4 c^2 (\lambda -1) S_{\phi }^2 \left(\xi  (\xi +1) (\lambda -\xi ) S_{3 \psi } S_{\phi
   }^2 \right. \\ 
   &\qquad \left. -S_{\psi } \left(\xi  C_{\phi }^2 (\lambda  (5-3 \xi )+(3-5 \xi ) \xi )+(\xi +1)
   \left(\lambda  \left((\xi -1) \xi ^2-1\right)+\xi ^3+\xi -1\right)\right)\right)
     \end{aligned}   
\ee

\end{document}